\documentclass[aps,pra,preprint,superscriptaddress]{revtex4}

\usepackage{color}
\usepackage{graphicx,epsfig,amssymb}
\usepackage{bm}

\begin{document}
\title{\bf Zero-energy bound state trapped in line-shaped vortex in topological superconductor}
\vspace{1.5em}
\author{P.~Schlottmann}
\affiliation{
Department of Physics, Florida State University, Tallahassee, Florida 32306}
\date{today}

\begin{abstract}
Fermion bound states in the core of a line-shaped vortex of a two-dimensional topological superconductor are investigated.  The superconducting pairing potential, described in terms of elliptical coordinates, vanishes along a line defect with the two foci at the endpoints. The superconductivity is induced into a topological insulator via proximity effect with a type II $s$-wave superconductor. The spin and the momentum are perpendicularly locked by the strong spin-orbit coupling via Rashba interaction. A zero-energy Majorana state arises from the Berry phase together with a sequence of equally spaced fermion excitations. By solving the Bogoliubov-de Gennes equations using the method employed by Caroli, de Gennes and Matricon we calculate the energies, the wave-functions and spin-polarization of the bound states. An analytic expression for the local density of states within the vortex is obtained.

\end{abstract}
\maketitle

\section{Introduction}

Majorana bound states are unconventional zero-energy quasi-particles with non-Abelian statistics which are their own antiparticles. \cite{Beenakker1} These states have been proposed for storing quantum information \cite{Kitaev} and for fault-tolerant quantum computing. The formation of Majorana bound states at the surface of a strong topological insulator (TI) with superconducting properties in the proximity of an $s$-wave superconductor (S) has been explored by Fu and Kane \cite{FuKane1,FuKane2} and Sau {\it et al.} \cite{Sau1} Following Refs. \cite{FuKane1} and \cite{FuKane2} the surface states of a heterostructure in a TI with proximity-induced S were studied in different geometries. \cite{Sau1,Chiu,Rakhmanov,Akzyanov1} The magnitude of the induced superconducting gap depends on the transparency of the TI/S junction and is in general smaller than the parent S gap. For low-lying excitations, the induced gap is a weak function of energy so that $\Delta$ can be considered constant. Accordingly the problem is reduced to a 2D topological electron gas with superconducting BCS gap. An alternative platform to generate Majorana zero-energy modes in heterostructures consists of a semiconducting thin film sandwiched between a magnetic insulator and an $s$-wave superconductor. \cite{Sau2,MaoZhang} In summary, a Majorana state can arise from a superconducting flux quantum trapped by a defect in a superconducting layer (Abrikosov vortex) \cite{Beenakker1,Berthod,Schl1,Schl2,Beenakker2}, a layered heterostructure of a magnetic insulator, a semiconductor and a superconductor \cite{Sau2,MaoZhang}, in a tri-junction pair geometry \cite{FuKane1,Sau1} or a cylindrical cavity in a 2D superconductor. \cite{Rakhmanov,Akzyanov1,Hassler} 

In this paper we investigate a flux quantum trapped in a line-shaped vortex core in a 2D topological superconductor. As a consequence of the geometry, all previously studied cases require polar coordinates, while in the present situation elliptic coordinates are more appropriate.

The electronic structure of in-gap states of a vortex in a 2D topological superconductor has been studied by numerous authors \cite{Sau1,Rakhmanov,Akzyanov1,Hassler,Ioselevich,Suzuki,Akzyanov2,Akzyanov3,Jackiw} by solving the corresponding Bogoliubov-de Gennes (BdG) equations. Key is the strong spin-orbit coupling leading to spin-momentum locking and a Berry phase of 1/2. This converts half-integer quantum numbers into integer ones and opens the possibility to create a zero-energy Majorana fermion. For a finite Fermi energy the low-energy excitations are equally spaced by an amount $\Delta_{\infty}^2/E_F$, where $\Delta_{\infty}$ is the superconducting gap far away from the vortex.  The minigap from the Majorana mode to the first excitation can however be tuned by varying the Fermi energy and for $E_F \to 0$ the gap is of the order of $\Delta_{\infty}$. \cite{Sau1,Rakhmanov,Schl1}

Vortex states in superconducting graphene \cite{Khaymovich} are closely related to vortices with pointlike core in the present problem. The pseudospin parametrizing the two sublattices of the honeycomb lattice plays the role of the spin in the topological insulator.  The BdG equations in superconducting graphene for energies close to the Dirac points (considering the spin, pseudospin and particle-holes) reduce to eight equations which decouple into two equivalent subsets of four equations each. Each subset is then equivalent to the present problem for a vortex trapped by a point defect with Dirac interaction. \cite{Beenakker3} 

Analytical expressions for the low-energy bound state eigenvalues, the eigenfunctions and the local density of states (LDOS) for an isolated point vortex with strong spin-orbit coupling have been recently obtained \cite{Schl1,Schl2,Hassler} by solving the BdG equations using the method employed by Caroli, de Gennes and Matricon (CdeGM) \cite{Caroli1,Caroli2} for a topological type II $s$-wave superconductor in terms of Bessel functions. The method consists of solving the BdG equations for small distances (compared to the correlation length $\xi$) from the core of the vortex, as well as for larger distances (still smaller than $\xi$). If the matching condition of these two solutions at an intermediate distance is independent of the distance from the vortex core, then we have a solution for the entire region of the vortex. 

Experimentally, systems with large superconducting transition temperature are desirable. A zero-energy mode was detected via spin-selective Andreev reflections in the heterostructure Bi$_2$Te$_3$/NbSe$_2$ \cite{Sun} and with tunneling scanning spectroscopy in monolayers of the high-temperature superconductor FeTe$_{0.55}$Se$_{0.45}$. \cite{Wang,Machida} To prove that the measured peak corresponds to a Majorana state requires the exchange of two such modes (qubit).

In the present paper we extend the calculation to a line-shaped vortex core (corresponding to a cut in the 2D superconductor confined between two foci) using elliptical coordinates rather than polar coordinates. The superconducting gap then vanishes along the entire line defect. In Sect. II we introduce the Rashba model and the BdG equations in elliptical coordinates, as well as the solution for the zero-energy Majorana state. The linear vortex traps one flux quantum. The general solution of the BdG equations is presented in the Appendix. Parametrizing the elliptical coordinates the problem is very similar to the solution of the BdG equations in spherical coordinates. Results are presented in Sect. III and Conclusions follow in Sect. IV.

Atomic line defects with zero-energy end-states have been observed in monolayer FeTe$_{0.5}$Se$_{0.5}$ high-$T_c$ superconductors. \cite{Chen} This situation is different from the one treated here, where the cut in the superconducting layer joins the two end-points. The superconducting phase field winds around the position of each vortex, i.e. once around the linear cut in the present case, while once around the positions of each of the vortex cores in \cite{Chen}.

The details of the calculation have been deferred to the Appendix, while the main results are presented in sections II, III and IV. It should be possible for the reader to skip the Appendix.  
 
\section{Rashba interaction}
\subsection{The model}

The 2D electron gas represents the surface states of a 3D topological insulator.  Superconductivity is induced via proximity by a type II $s$-wave superconductor. \cite{Sau1,
Zyuzin,Bobkova} The Rashba spin-orbit interaction locks the spin perpendicular to the momentum and the magnetic field is oriented perpendicular to the plane of the surface.

The wave function is a 4-component spinor, $\Psi({\bf r}) = [\psi_{\uparrow}({\bf r}) \ \psi_{\downarrow}({\bf r}) \ \psi^{\dagger}_{\uparrow}({\bf r}) \ \psi^{\dagger}_{\downarrow}({\bf r})]^T$, and the Hamiltonian is given by ${\cal H} = {\textstyle \frac{1}{2}} \int d^2r \Psi^{\dagger}({\bf r}) {\check {\cal H}}_B^{\perp}({\bf r}) \Psi({\bf r})$, where
\begin{eqnarray}
&&{\check {\cal H}}_B^{\perp}({\bf r}) = \left[ \begin{array}{cc} {\hat h}({\bf r}) & {\hat \Delta({\bf r})} \\
- {\hat \Delta^*({\bf r})} & - {\hat h}^*({\bf r}) \end{array} \right] \ \ , \label{HB} \\
&&{\hat h}({\bf r}) = v_F {\boldsymbol{\hat \sigma}} \cdot \Bigl[\Bigl({\bf p} - \frac{e}{c} {\bf A({\bf r})} \Bigr)\times {\bf e}_z\Bigr] - E_F{\hat I} \ \ , \label{h} \\
&&{\hat \Delta}({\bf r}) = \Delta({\bf r}) i {\hat \sigma}_y \ \ , \label{Delta}
\end{eqnarray}
where ${\bf e}_z$ is the vector normal to the plane, $v_F$ the velocity in the Dirac cone, ${\hat I}$ the unit matrix and ${\bf A({\bf r})}$ the vector potential. Using similar arguments as in references \cite{Caroli1,Caroli2} the vector potential and the magnetic field can be neglected in Eq. (\ref{h}).

\subsection{Elliptic coordinates}

Since in this paper we consider a linearly shaped vortex, i.e. a linear cut in the 2D superconductor, it is appropriate to employ elliptical coordinates, \cite{wiki} rather than the usual polar ones. Elliptic coordinates are a 2D orthogonal system in which the coordinate lines are confocal ellipses and hyperbolae. The two foci $F_1$ and $F_2$ are taken at $\pm a$ on the $x$-axis of the Cartesian coordinate system.

We define the elliptic coordinates $(\mu,\nu)$ as (see Fig. [1])
\begin{equation}
x = a \cosh(\mu) \cos(\nu) \ \ \ \ \ , \ \ \ \ \ y = a \sinh(\mu) \sin(\nu) \ \ , \label{elliptic1}
\end{equation} 
where $\mu$ is a nonnegative real number and $\nu$ belongs to the interval $[0,2\pi )$. This can be rewritten on the complex plane as
\begin{equation}
X_{\pm} = x \pm iy = a \cosh(\mu \pm i\nu) \ \ . \label{elliptic2}
\end{equation}

\vskip 0.15in
\begin{figure}[!ht]
\begin{center}
\resizebox{0.45\textwidth}{!}
{\includegraphics{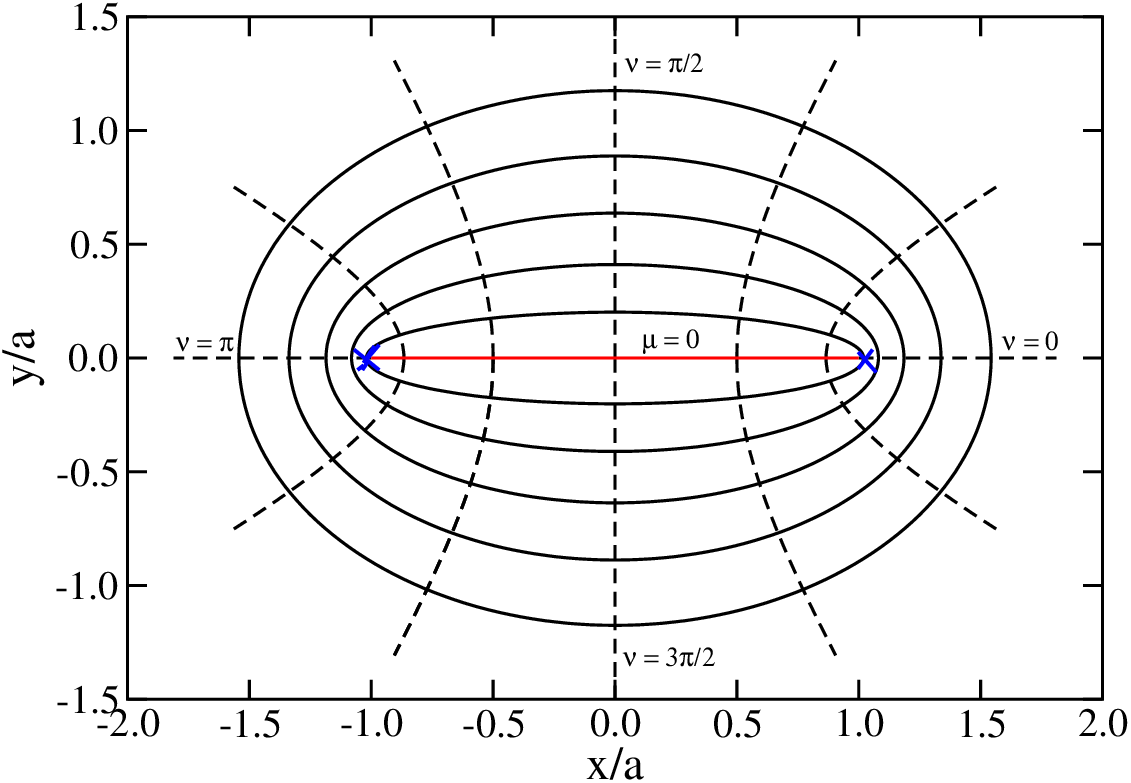}}
\end{center}
\caption{({\it color online}) Ellipses and hyperbolae with foci at $x=\pm a$ indicated by the blue crosses. The red straight line joining the two foci corresponds to $\mu=0$. The ellipses are parametrized by $\mu=0.2, 0.4, 0.6, 0.8, 1.0$ (solid curves), while the hyperbolae by $\nu=0,$ $\pi/6, \pi/3, \pi/2, 2\pi/3, 5\pi/6, \pi, 7\pi/6, 4\pi/3, 3\pi/2, 5\pi/3, 11\pi/6$ (dashed curves). The red segment denotes the cut along which the superconducting gap vanishes. The gap gradually closes with distance from the linear cut, i.e. as a function of increasing $\mu > 0$ or the size of the ellipse. }
\label{ellipses}
\end{figure}

These definitions correspond to ellipses and hyperbolae for constant $\mu$ and $\nu$, respectively, i.e.
\begin{eqnarray}
\frac{x^2}{a^2 \cosh^2(\mu)} + \frac{y^2}{a^2 \sinh^2(\mu)} &=& 1 \ \ , \nonumber \\
\frac{x^2}{a^2 \cos^2(\nu)} - \frac{y^2}{a^2 \sin^2(\nu)}&=& 1 \ \ . \label{ell-hyp}
\end{eqnarray}

An alternative definition of elliptical coordinates is $(\sigma,\tau)$, where $\sigma = \cosh(\mu)$ and $\tau = \cos(\nu)$. The curves of constant $\sigma$ are then ellipses and those of constant $\tau$ hyperbolae. Hence, the coordinate $\tau$ belongs to the interval $[-1,1]$, whereas $\sigma$ must be greater than or equal to one. There is a simple geometrical relation of the distances to the foci $F_1$ and $F_2$. For any point in the plane, the sum $r_1 + r_2$ of its distances to the foci equals $2a\sigma$, whereas their difference $r_1-r_2$ equals $2a\tau$. Thus, the distance to $F_1$ is $a(\sigma + \tau$), and the distance to $F_2$ is $a(\sigma - \tau)$. 
 
According to Abrikosov \cite{Abrikosov} the radial dependence of an isolated superconducting flux point vortex is approximately $\Delta(r) = \Delta_{\infty}\tanh(r/\xi)$. For a distance $r$ much shorter than the coherence length $\xi$, $\Delta(r)$ is linear in $r$, if the vorticity is one. The sum of the distances to both foci is then $r_1+r_2 = 2a\sigma$ and due to the linearity in $r$, the sum of the pair potential for both vortices is $\Delta(r_1)+\Delta(r_2)-\Delta(2a) = \Delta[2a(\sigma-1)]$, i.e. it does not depend on $\tau$ and vanishes for $\mu = 0$.

A sample is prepared by depositing an $s$-wave superconductor with bulk pairing potential $\Delta_{SC}$ on the surface of a 3D topological insulator. The proximity of the two materials induces a pairing potential $\Delta_{TI}$ on the surface of the $TI$. The function  $\Delta_{TI}({\vec r})$ depends on the transparency of the junction, but ideally the two pairing functions are proportional, so that the subindex in $\Delta_{TI}({\vec r})$ can be dropped.

A point defect at the interface can pin an Abrikosov vortex. However, more complex vortex patterns, such as a line defect, requires a more elaborate manufacturing technique, e.g. vapor deposition with a shadow mask or a photo-lithographic structuring or molecular beam epitaxy. The straight segment is schematically shown by the red line in Fig. 1 ($\mu=0$). The superconducting gap vanishes by construction along the segment.  The gap closes gradually as a function of the distance from the line. 

\subsection{BdG equations in elliptic coordinates}

The momentum can be expressed in terms of the derivatives with respect to $X_{\pm}$, i.e.
\begin{equation} 
\frac{\partial}{\partial X_{\pm}} = - \frac{i}{a \sinh(\mu \pm i\nu)} \left[ i \frac{\partial}{\partial \mu} \pm \frac{\partial}{\partial \nu} \right] \ \ , \label{deriv}
\end{equation}
so that ${\hat h}({\bf r}) = $ 
\begin{eqnarray}
 \left[ \begin{array}{ll} -E_F & - \frac{i v_F}{a \sinh(\mu + i\nu)} \Bigl( i \frac{\partial}{\partial \mu} + \frac{\partial}{\partial \nu} \Bigr) \\
\frac{i v_F}{a \sinh(\mu - i\nu)} \Bigl( i \frac{\partial}{\partial \mu} - \frac{\partial}{\partial \nu} \Bigr) & -E_F \end{array} \right]  . \label{hh}
\end{eqnarray}
If only one flux quantum is contained in the vortex, we may write ${\hat \Delta(\bf r)} = \Delta[2a(\sigma-1)]i {\hat \sigma}_y e^{-i\nu}$. The $e^{-i\nu}$ phase of ${\hat \Delta(\bf r)}$ can be eliminated via a gauge transformation. 

The field operators are expanded as 
\begin{equation}
\psi(\mu,\nu) = \frac{1}{\sqrt{2\pi}} \sum_m \psi_m(\mu) e^{im\nu} \ , \label{fieldop}
\end{equation}
\noindent where $m$ is an integer for vorticity one and similarly  
\begin{equation}
\frac{1}{\sinh(\mu \pm i\nu)} = 2 \sum_{n=0}^{\infty} \exp[-(2n+1)\mu \mp i(2n+1)\nu] \ \ . \label{sinh}
\end{equation}
The spinor $\Psi$ is required to be a single-valued function of $\nu$, so that
\begin{eqnarray}
\Psi_m(\mu,\nu) = \left[ \begin{array}{c} \psi_{\uparrow m}(\mu) e^{-i\nu} \\ \psi_{\downarrow m}(\mu) \\ \psi_{\uparrow m}^{\dagger}(\mu) e^{i\nu} \\ \psi_{\downarrow m}^{\dagger}(\mu) \end{array} \right] \ \ . \label{Psi}
\end{eqnarray}
Applying ${\hat h}(\bf r)$ to the spinor we obtain
\begin{eqnarray}
\left[ \begin{array}{cc} -E_F & \frac{v_F}{a \sinh(\mu + i\nu)} \bigl( \frac{\partial}{\partial \mu} + m \bigr) \\
- \frac{v_F}{a \sinh(\mu - i\nu)} \bigl(\frac{\partial}{\partial \mu} -m + 1 \bigr) & -E_F \end{array} \right] 
\left[ \begin{array}{cc}
\psi_{\uparrow m}(\mu) e^{i(m-1)\nu} \\ \psi_{\downarrow m}(\mu) e^{im\nu} \end{array} \right] . \label{h1}
\end{eqnarray}
and
\begin{eqnarray}
\left[ \begin{array}{cc} E_F & -\frac{v_F}{a \sinh(\mu - i\nu)} \bigl( \frac{\partial}{\partial \mu} + m \bigr) \\
\frac{v_F}{a \sinh(\mu + i\nu)} \bigl(\frac{\partial}{\partial \mu} -m - 1 \bigr) & E_F \end{array} \right] 
\left[ \begin{array}{cc}
\psi_{\uparrow m}^{\dagger}(\mu) e^{i(m+1)\nu} \\ \psi_{\downarrow m}^{\dagger}(\mu) e^{im\nu} \end{array} \right] . \label{h2}
\end{eqnarray}
Inserting ${\hat \Delta}(\bf r)$ and the Taylor expansion of $1/\sinh(\mu \pm i\nu)$ [Eq. (\ref{sinh})] the differential equation yields
\begin{eqnarray}
\sum_{n=0}^{\infty} &&\left[ \begin{array}{cccc} -(E_F+E)\delta_{n,0} & Z_n \bigl(\frac{\partial}{\partial \mu} + m \bigr) & 0 & {\overline \Delta} e^{-i\nu} \delta_{n,0} \\
-Z^*_n \bigl(\frac{\partial}{\partial \mu} - m+1 \bigr) & -(E_F+E)\delta_{n,0} &  -{\overline \Delta} e^{-i\nu} \delta_{n,0} & 0 \\
0 & {\overline \Delta} e^{i\nu} \delta_{n,0} & (E_F-E)\delta_{n,0} & - Z^*_n \bigl(\frac{\partial}{\partial \mu} + m \bigr) \\
-{\overline \Delta} e^{i\nu} \delta_{n,0} & 0 & Z_n \bigl(\frac{\partial}{\partial \mu} - m -1\bigr) & (E_F-E)\delta_{n,0} \end{array} \right] \nonumber
\end{eqnarray}
\begin{eqnarray}
\times \left[\begin{array}{c} \psi_{\uparrow m n}(\mu) e^{i(m-1)\nu} \\ \psi_{\downarrow m n}(\mu) e^{im\nu} \\ \psi_{\uparrow m n}^{\dagger}(\mu)e^{i(m+1)\nu} \\ \psi_{\downarrow m n}^{\dagger}(\mu) e^{im\nu} \end{array} \right] 
= \left[ \begin{array}{c} 0 \\ 0 \\ 0 \\ 0 \end{array} \right] , \label{matrix} 
\end{eqnarray} 
where $Z_n$ denotes $\frac{2v_F}{a}e^{-(2n+1)(\mu+i\nu)}$ and ${\overline \Delta} = \Delta[2a(\sigma-1)]$, with $\sigma$ being $\cosh(\mu)$. Here we have incorporated the subindex $n$ into the field operator $\psi$. It is easy to verify that $\psi_{\uparrow m n} = \psi_{\downarrow m n} = \psi_{\uparrow m n}^{\dagger} = \psi_{\downarrow m n}^{\dagger} = 0$ for $n \neq 0$. Only $n=0$ plays a relevant role and the subindex $n$ can be suppressed. Below we denote the components of the spinor $\Psi$ as $f_j^m(\mu)$ with $j=1,\cdots,4$, where the index $j$ contains the spin (up or down) and annihilation and creation symbols. The $\nu$-dependence of $f_j^m$ is $\exp[-i\nu {\hat \tau}_z (1+{\hat \sigma}_z/2 +im\nu]$, where ${\hat \sigma}_z$ and ${\hat \tau}_z$ are Pauli matrices for the spin and particle/hole sectors, respectively.

The matrix equation reduces to four linearly coupled differential equations, in which the $\nu$ dependence cancels out and the solution is single valued,
\begin{eqnarray}
&&\frac{2v_F}{a} e^{-\mu} \Bigl(\frac{\partial}{\partial \mu} - m+1 \Bigr)f^m_1(\mu) + (E+E_F)f^m_2(\mu) + {\overline \Delta} f^m_3(\mu) = 0 \ \ , \label{f1a} \\
&&\frac{2v_F}{a} e^{-\mu} \Bigl(\frac{\partial}{\partial \mu} + m \Bigr)f^m_2(\mu) - (E+E_F)f^m_1(\mu) + {\overline \Delta} f^m_4(\mu) = 0 \ \ , \label{f2a} \\
&&\frac{2v_F}{a} e^{-\mu} \Bigl(\frac{\partial}{\partial \mu} + m+1 \Bigr)f^m_3(\mu) - (E-E_F)f^m_4(\mu) + {\overline \Delta} f^m_1(\mu) = 0 \ \ , \label{f3a} \\
&&\frac{2v_F}{a} e^{-\mu} \Bigl(\frac{\partial}{\partial \mu} - m \Bigr)f^m_4(\mu) + (E-E_F)f^m_3(\mu) + {\overline \Delta} f^m_2(\mu) = 0 \ \ . \label{f4a}
\end{eqnarray}
Defining $\rho = e^{\mu} a/2$, $E_F=v_Fk_F$, ${\tilde \rho} = \rho k_F$, ${\tilde E} = E/E_F$ and ${\tilde \Delta} = {\overline \Delta}/E_F$, the equations are dimensionless and have the following form
\begin{eqnarray}
&&\Bigl(\frac{\partial}{\partial {\tilde \rho}} - \frac{m-1}{\tilde \rho} \Bigr)f^m_1({\tilde \rho}) + (1+{\tilde E})f^m_2({\tilde \rho}) + {\tilde \Delta} f^m_3({\tilde \rho}) = 0 , \label{f1b} \\
&&\Bigl(\frac{\partial}{\partial {\tilde \rho}} + \frac{m}{\tilde \rho} \Bigr)f^m_2({\tilde \rho}) - (1+{\tilde E})f^m_1({\tilde \rho}) + {\tilde \Delta} f^m_4({\tilde \rho}) = 0 , \label{f2b} \\
&&\Bigl(\frac{\partial}{\partial {\tilde \rho}} + \frac{m+1}{\tilde \rho} \Bigr)f^m_3({\tilde \rho}) + (1-{\tilde E})f^m_4({\tilde \rho}) + {\tilde \Delta} f^m_1({\tilde \rho}) = 0  , \label{f3b} \\
&&\Bigl(\frac{\partial}{\partial {\tilde \rho}} - \frac{m}{\tilde \rho} \Bigr)f^m_4({\tilde \rho}) - (1 -{\tilde E})f^m_3({\tilde \rho}) + {\tilde \Delta} f^m_2({\tilde \rho}) = 0  . \label{f4b}
\end{eqnarray}
These equations are formally similar to those of a single pinned point vortex \cite{Schl1,Rakhmanov,Akzyanov1} described by polar coordinates. However $\rho (=e^{\mu} a/2)$ is {\it not} the radial coordinate so that the meaning of these equations in Cartesian coordinates is substantially different. $\nu$ and $\theta$ are both angular coordinates.

\subsection{Majorana state}

The Majorana bound state has zero energy and corresponds to $m=0$. It is easily verified that the solution of Eqs. (\ref{f1b}) - (\ref{f4b}) is given by
\begin{eqnarray}
&&f^0_1({\tilde \rho}) = C J_1({\tilde \rho}) e^{-K({\tilde \rho})} \ , \nonumber \\
&&f^0_2({\tilde \rho}) = -C J_0({\tilde \rho}) e^{-K({\tilde \rho})} \ , \nonumber \\
&&f^0_3({\tilde \rho}) = C J_1({\tilde \rho}) e^{-K({\tilde \rho})} \ , \nonumber \\
&&f^0_4({\tilde \rho}) = -C J_0({\tilde \rho}) e^{-K({\tilde \rho})} \ , \label{Majorana}
\end{eqnarray}
where $J_n$ are Bessel functions of integer order, $C$ is a normalization constant and $K({\tilde \rho})$ is given by
\begin{eqnarray}
&&\frac{d K}{d{\tilde \rho}} = \frac{\Delta[2a(\sigma-1)]}{E_F} \ , \nonumber \\
&&K({\tilde \rho}) = \int^{\tilde \rho}_{ak_F/2} d{\tilde \rho'}\Delta\Bigl[\frac{2{\tilde \rho'}}{k_F} + \frac{a^2k_F}{2{\tilde \rho'}} - 2a \Bigr]/E_F \ . \label{K}
\end{eqnarray}
Note that the definition of $K({\tilde \rho})$ is different from the corresponding expression for a point vortex in polar coordinates. \cite{Schl1,Schl2} The Majorana wave function is
\begin{eqnarray}
\Psi_M &=& C \int d^2r e^{-K({\tilde \rho})} \Bigl[ J_1({\tilde \rho}) e^{-i\nu} \psi_{\uparrow}({\bf r}) - J_0({\tilde \rho}) \psi_{\downarrow}({\bf r}) \nonumber \\
&+& J_1({\tilde \rho}) e^{i\nu} \psi^{\dagger}_{\uparrow}({\bf r}) - J_0({\tilde \rho}) \psi_{\downarrow}^{\dagger}({\bf r}) \Bigr] \ . \label{maj}
\end{eqnarray}

A Majorana state is self-adjoint and should satisfy $\Psi_M=\Psi_M^{\dagger}$. The adjoint of the first term in (\ref{maj}) yields the third term and the adjoint of the second term yields the fourth term.

\subsection{Energies and wave functions}

The solution of the BdG equations is outlined in the Appendix. The energy of the $m^{th}$ bound state is given by
\begin{equation}
E_m = m \int_{{\tilde \rho}_c}^{\infty} dx \exp[-2K(x)] \frac{\tilde \Delta}{x} \bigm/ \int_{{\tilde \rho}_c}^{\infty} dx \exp[-2K(x)] . \label{Em} 
\end{equation}
The excited states are then equally spaced energy levels.

Approximate expressions to leading order for the amplitudes of the wave functions for the excited bound states, $m \ge 1$, are given by 
\begin{eqnarray}
&&f_1^m({\bf r}) = C' J_{m-1}({\tilde \rho}) e^{-K({\tilde \rho})} e^{i(m-1)\nu} , \nonumber \\
&&f_2^m({\bf r}) = C' J_m({\tilde \rho}) e^{-K({\tilde \rho})} e^{im\nu} , \nonumber \\
&&f_3^m({\bf r}) = -C' J_{m+1}({\tilde \rho}) e^{-K({\tilde \rho})} e^{i(m+1)\nu} , \nonumber \\
&&f_4^m({\bf r}) = C' J_m({\tilde \rho}) e^{-K({\tilde \rho})} e^{im\nu} , \label{wave}
\end{eqnarray}
and the energy wave function with energy $E_m$ is then
\begin{eqnarray}
{\hat \psi}_E &=& C' \int d^2r e^{-K({\tilde \rho})} \Bigl[J_{m-1}({\tilde \rho}) e^{-i\nu} \psi_{\uparrow}({\bf r}) + J_m({\tilde \rho}) \psi_{\downarrow}({\bf r}) \nonumber \\
&&- J_{m+1}({\tilde \rho}) e^{i\nu} \psi_{\uparrow}^{\dagger}({\bf r}) + J_m({\tilde \rho}) e^{i\nu} \psi_{\downarrow}^{\dagger}({\bf r})\Bigr] e^{im\nu} . \label{WF}
\end{eqnarray}
For $m \neq 0$ it represents an ordinary fermion wave function (not self-adjoint).
 
\section{Results}

\subsection{Excitation energies}

The solution of the BdG equations for the general case is outlined in the Appendix, where we have shown that the wave functions of the fermion excited states consist of two factors, 

\begin{figure}[!ht]
\begin{center}
\resizebox{0.50\textwidth}{!}
{\includegraphics{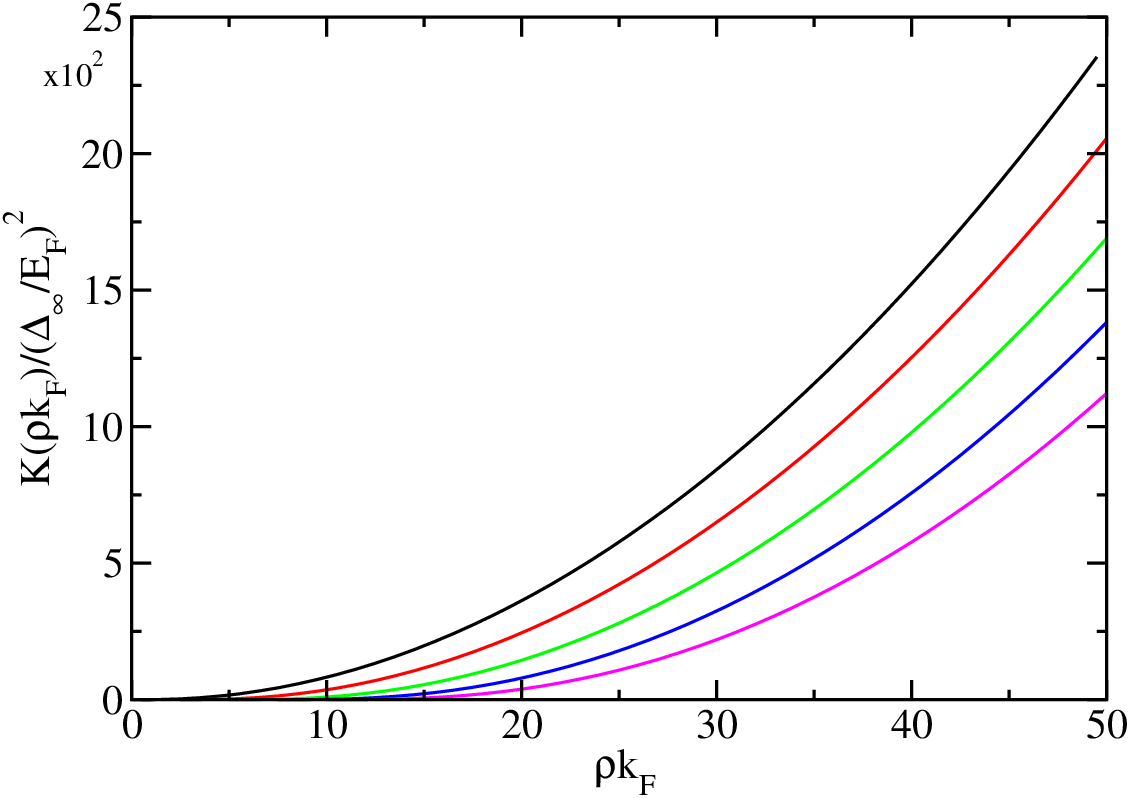}}
\end{center}
\caption{({\it color online}) The function $K({\tilde \rho})$ normalized to $(\Delta_{\infty}/E_F)^2$ for five values of $a k_F$, namely 1 (black), 5 (red), 10 (green), 15 (blue), and 20 (magenta). At large distances the wave function falls off faster with increasing $K$. $K({\tilde \rho})$ is approximately a parabolic function.}
\vskip 0.30in
\label{FigAK}
\end{figure}

\par\noindent one being a Bessel function and the second one determines the exponential decay of the wave function. The latter, $\exp(-K({\tilde \rho}))$, is a function of the superconducting order parameter. The energies are equally spaced and depend on the function $K({\tilde \rho})$, Eq. (\ref{K}). Since the superconducting gap is a linear function of distance 
with slope $\Delta_{\infty}/\xi$, where $\xi$ is the coherence length, $K({\tilde \rho})$ is easily evaluated yielding
\begin{eqnarray}
K({\tilde \rho}) &=& \Bigl(\frac{\Delta_{\infty}}{E_F}\Bigr)^2 \Bigl[{\tilde \rho}^2 -2ak_F{\tilde \rho} \nonumber \\
&&+\frac{(ak_F)^2}{2} \Bigl(\ln(2 {\tilde \rho}/ak_F) + \frac{3}{2} \Bigr) \Bigr] \ . \label{KK1}
\end{eqnarray}
As shown in Fig. [\ref{FigAK}] the function $K({\tilde \rho})$ is approximately parabolic and decreases with increasing the distance between the foci $2a$.

\vskip 0.20in
\begin{figure}[!ht]
\begin{center}
\resizebox{0.50\textwidth}{!}
{\includegraphics{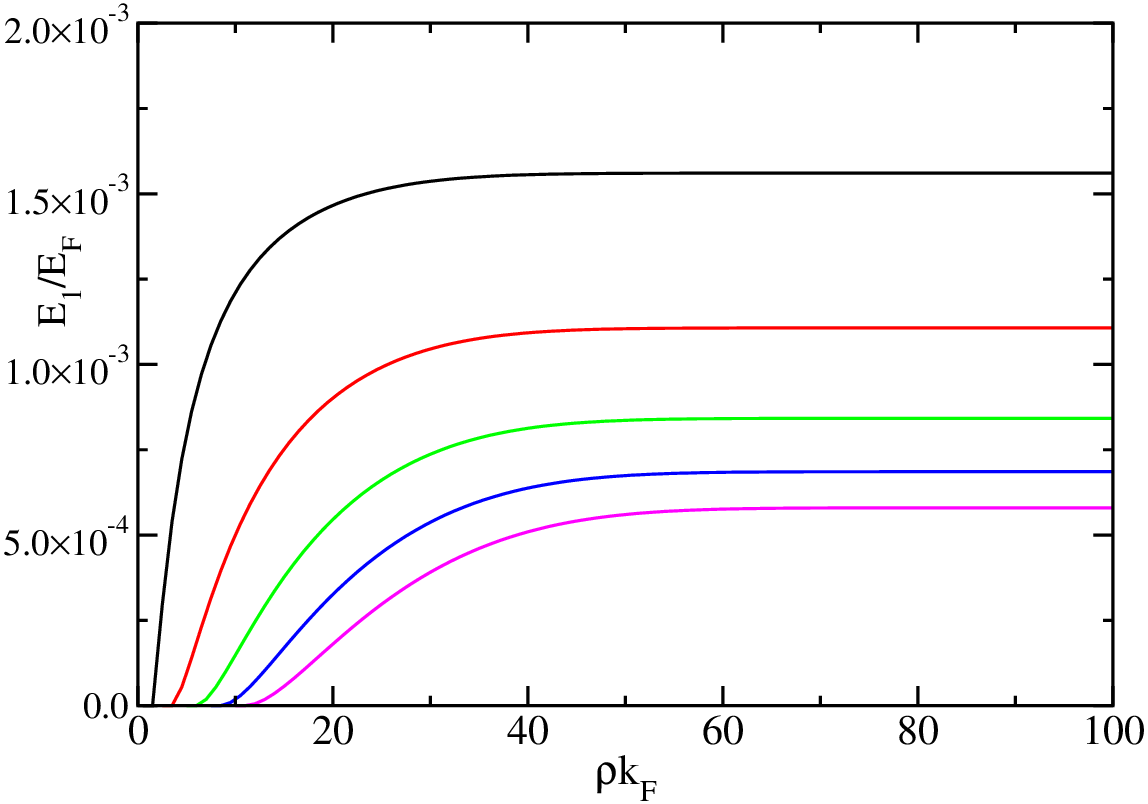}}
\end{center}
\caption{({\it color online}) First excited state energy $E_1$: In Eq. (\ref{Em}) we set ${\tilde \rho}_c = ak_F/2$ and replace the upper integration limit $\infty$ by ${\tilde \rho}$. For large ${\tilde \rho}$, $E_1$ saturates yielding the excited energy. Five values of $ak_F$ are considered, 1 (black), 5 (red), 10 (green), 15 (blue), and 20 (magenta), where $2a$ is the distance between the foci. The energy of the excited state decreases with increasing $a k_F$.} 
\label{FigWW}
\end{figure}
The energies are given by Eq. (\ref{Em}), where the matching point ${\tilde \rho}_c$ at the lower integration limit can be approximated by $ak_F/2$. It needs to be verified that as a function of the upper integration limit the value of the $E_m$ converges. This convergence is shown in Fig. [\ref{FigWW}] for the same values of $a k_F$ as in Fig. [\ref{FigAK}]. The energy saturates into plateaus with increasing ${\tilde \rho}$. The gap between consecutive excited states decreases with increasing $a k_F$. For given $ak_F$ the excited states are equidistant, i.e. $E_m = m E_1$, where $m=0$ corresponds to the Majorana state.

\subsection{Local density of bound states}

The one-particle Green's function for spin-component $\sigma$ is given by the wave functions and energy eigenvalues \cite{Bardeen}
\begin{equation}
G_{\omega}^{\sigma}({\bf r},{\bf r}') = \sum_m \left( \frac{u^{\sigma}_m({\bf r})u^{\sigma *}_m({\bf r})'}{\omega + i0 - E_m} + \frac{v^{\sigma}_m({\bf r})v^{\sigma *}_m({\bf r})'}{\omega + i0 + E_m}\right) , \label{Green}
\end{equation}
where $u$ and $v$ correspond to the functions $f^m_j({\bf r})$, i.e. the wave functions for particles and holes. The local density of states (LDOS) for spin projection $\sigma$ is given by the imaginary part of the Green's function for ${\bf r}' \to {\bf r}$
\begin{equation}
N_{\sigma}(\omega,{\tilde \rho}) = \sum_m N_{m \sigma}(\omega,{\tilde \rho}) = \sum_m \Bigl(|u^{\sigma}_m({\tilde \rho})|^2\delta(\omega-E_m)+|v^{\sigma}_m({\tilde \rho})|^2\delta(\omega+E_m)\Bigr) \ .  \label{Nomega}
\end{equation}
At finite temperature the $\delta$-function is substituted by minus the derivative of the Fermi function, and hence the peaks broaden with increasing temperature. \cite{Suzuki} 

The LDOS for our model with Rashba interaction is proportional to (up to a normalization constant for the spinor)
\begin{eqnarray}
N_{\uparrow}(\omega,{\tilde \rho}) &\propto& \sum_m \Bigl[J_{m-1}({\tilde \rho})^2 \delta(\omega-E_m) + J_{m+1}({\tilde \rho})^2 \delta(\omega+E_m)\Bigr] e^{-2 K({\tilde \rho})} \ , \nonumber \\
N_{\downarrow}(\omega,{\tilde \rho}) &\propto& \sum_m \Bigl[J_m({\tilde \rho})^2 \delta(\omega-E_m) + J_m({\tilde \rho})^2 \delta(\omega+E_m)\Bigr] e^{-2 K({\tilde \rho})} \ . \label{Nsigma}
\end{eqnarray}

The above expressions are formally identical to the ones for a point vortex with two important differences, namely, the function $K({\tilde \rho})$ is not the same and in the present case ${\tilde \rho}$ is not the distance to the origin of the vortex, but a function of elliptical coordinates. The zero energy Majorana state corresponds to $m=0$ and $E_0=0$. The total LDOS ($N=N_{\uparrow}+N_{\downarrow}$) is invariant under the transformation $m \to -m$, i.e. $E_m \to -E_m$, and $u \leftrightarrow v$, i.e. the simultaneous interchange of the particle and hole amplitudes. It is interesting to notice that as for a point vortex the LDOS is not particle-hole symmetric. 

To understand the lack of particle-hole symmetry we consider Eqs. (\ref{h1}) and (\ref{h2}) and arguments similar to Ref. \cite{Schl1}. For down-spin electrons the effective angular momentum is $m$, while for up-spin electrons it is $(m-1)$. On the other hand, for down-spin holes it is still $m$, but for up-spin holes it is $(m+1)$. As a consequence of the difference between electrons and holes with up-spin, the particle-hole symmetry is broken. 

\begin{figure*}[!ht]
\begin{center}
\resizebox{0.45\textwidth}{!}
{\includegraphics{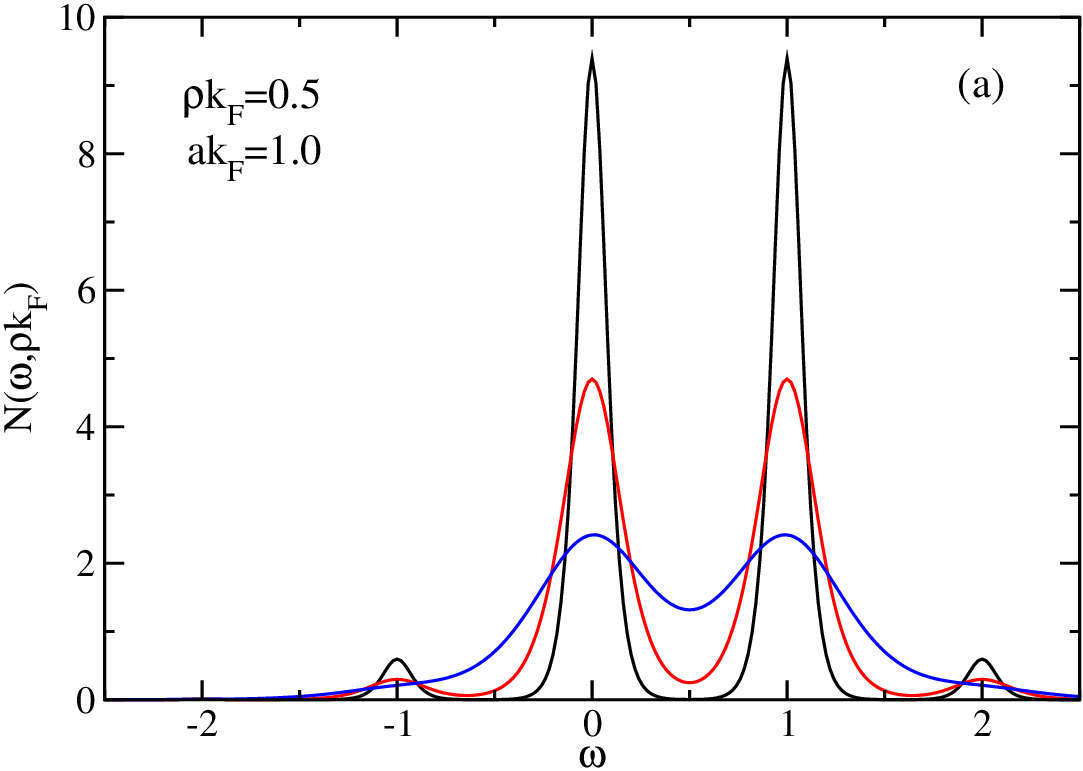}}
\hskip 0.15in
\resizebox{0.45\textwidth}{!}
{\includegraphics{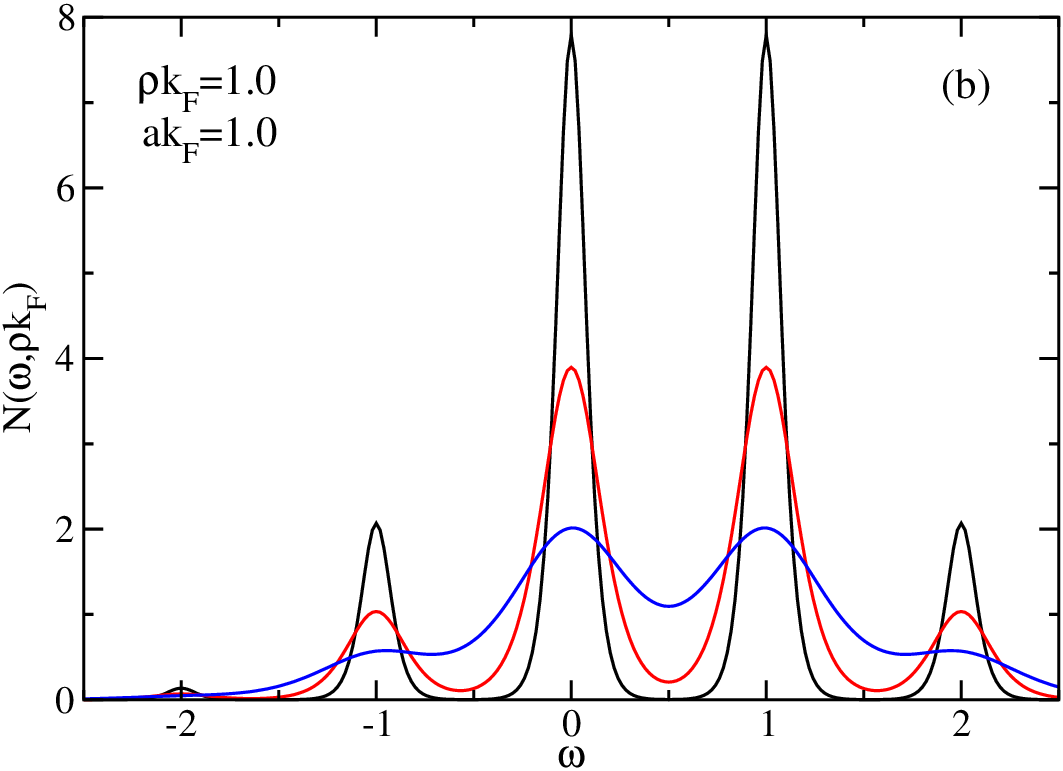}}
\vskip 0.35in
\resizebox{0.45\textwidth}{!}
{\includegraphics{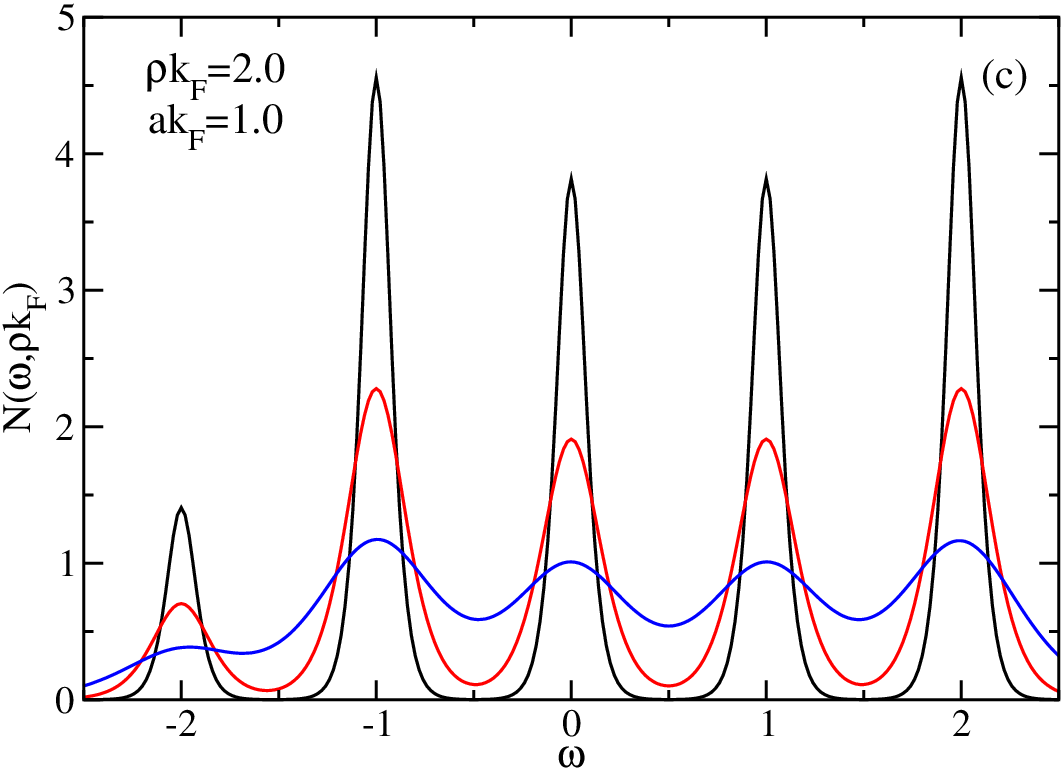}}
\end{center}
\caption{({\it color online}) LDOS of the bound states in the linear vortex of length $2a$ as a function of energy $\omega$ for $ak_F=1.0$ and three values of ${\tilde \rho}$, namely, 0.5 in panel (a), 1.0 in panel (b) and 2.0 in panel (c). The three curves in each panel represent different temperatures: $T=0.05$ (black), $T=0.1$ (red) and $T=0.2$ (blue).  In the limit $T \to 0$ the peaks are delta-functions.  The function $K$ is given by Eq. (\ref{KK1}). All energies are given in units of $E_1$, i.e. the spacing between excitations, and the LDOS is in arbitrary units but the same units for all three panels. Note that the LDOS is not particle-hole symmetric. Since ${\tilde \rho} \ge ak_F/2$ more than two peaks have nonzero spectral weight.}
\label{LDOS}
\end{figure*}

\begin{figure*}[!ht]
\begin{center}
\vskip 0.2in
\resizebox{0.45\textwidth}{!}
{\includegraphics{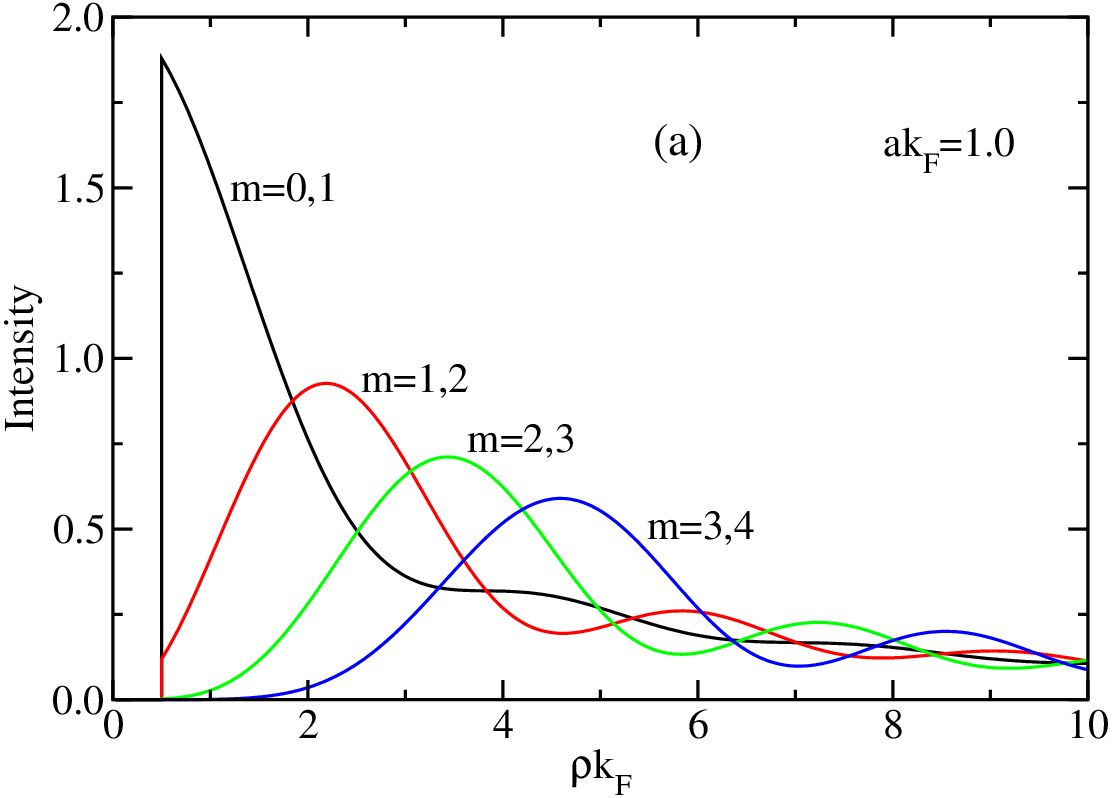}}
\hskip 0.34in
\resizebox{0.45\textwidth}{!}
{\includegraphics{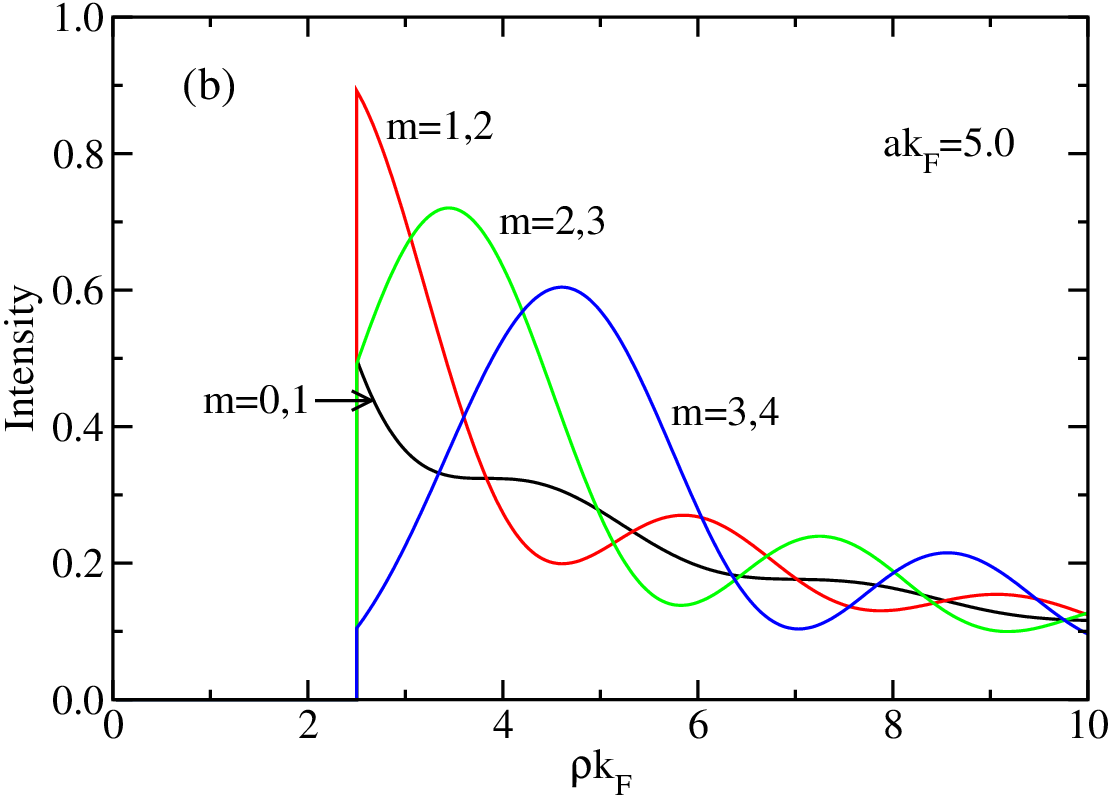}}
\end{center}
\caption{({\it color online}) Intensities of the peaks as a function of ${\tilde \rho}$ for several $m$. In general the intensity decreases with ${\tilde \rho}$, $m$ and $ak_F$, but not necessarily monotonically. $ak_F=1$ for panel (a) and $ak_F=5$ for panel (b). Note that ${\tilde \rho} \ge ak_F/2$. The colors of the curves are: $m=0,1$ (black), $m=1,2$ (red), $m=2,3$ (green) and $m=3,4$ (blue).}
\label{NN}
\end{figure*}

The LDOS is measurable via scanning tunneling microscopy (STM) by fine-tuning the energy $\omega$ at a distance from the core of the linear vortex, parametrized by ${\tilde \rho}$.

The LDOS as a function of energy in units of the gap $E_1$ is shown in Fig. (\ref{LDOS}) for three values of ${\tilde \rho}$. The intensity of the peaks depends on (i) the Bessel functions, (ii) the exponential decay due to $\exp[-K({\tilde \rho})]$, and (iii) on the separation of the two foci, $2a$. The LDOS also decreases rapidly as a function of temperature. The Majorana state ($m=0$) has always the same intensity as the $m=1$ peak. In general, the LDOS is symmetric about $\omega=0.5$. For ${\tilde \rho}=0$ and zero distance between the foci, $a=0$, i.e. for a simple point vortex, there are only two peaks, namely $m=0$ and 1, as a consequence of the Bessel functions, $J_m({\tilde \rho})$, which are zero as ${\tilde \rho} \to 0$ for $m \neq 0$. More peaks appear at finite ${\tilde \rho}$ and/or finite $ak_F$, but for larger ${\tilde \rho}$ or $ak_F$ the intensity does not necessarily decrease monotonically with $m$ as a consequence of the oscillations of the Bessel functions. 

For example in Fig. (4c) for $ak_F=1.0$ and ${\tilde \rho} =2$ the $m=2$ peak has higher intensity than the $m=1$ state. This can be understood with the plot of the intensities as a function of ${\tilde \rho}$ displayed in Fig. (\ref{NN}a). For ${\tilde \rho}=2$ the $m=0$ and 1 peaks have smaller intensity than the $m=2$ and 3 peaks. But for larger energies the intensity of the peaks decreases rapidly. For $a=0$ the main maxima of the LDOS form concentric circles about the center of the core with their radius increasing with $m$. For $a \neq 0$, on the other hand, the main maxima of the LDOS follow ellipses. 

The intensities of the peaks are shown in Fig. (\ref{NN}) as a function of ${\tilde \rho}$ and for two values of $ak_F$. Note that ${\tilde \rho} \ge ak_F/2$.  The peaks are denoted by two consecutive $m$ values indicating, e.g. $[J_m({\tilde \rho})^2 + J_{m+1}({\tilde \rho})^2]\exp(-2K{\tilde \rho})$; the $m=0,1$ curve is black, the $m=1,2$ curve is red, the $m=2,3$ curve is green and the $m=3,4$ curve is blue. As already mentioned, the oscillations arise from the Bessel functions. 

\subsection{Spin polarization}

\begin{figure*}[!ht]
\begin{center}
\resizebox{0.45\textwidth}{!}
{\includegraphics{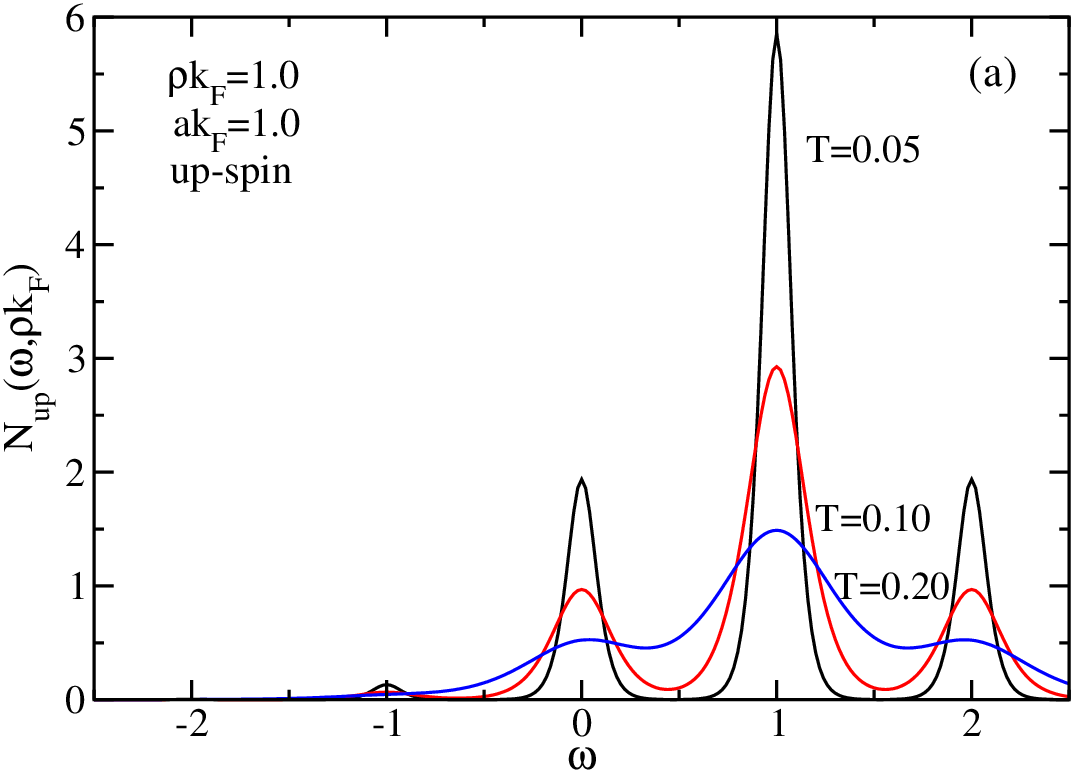}}
\hskip 0.34in
\resizebox{0.45\textwidth}{!}
{\includegraphics{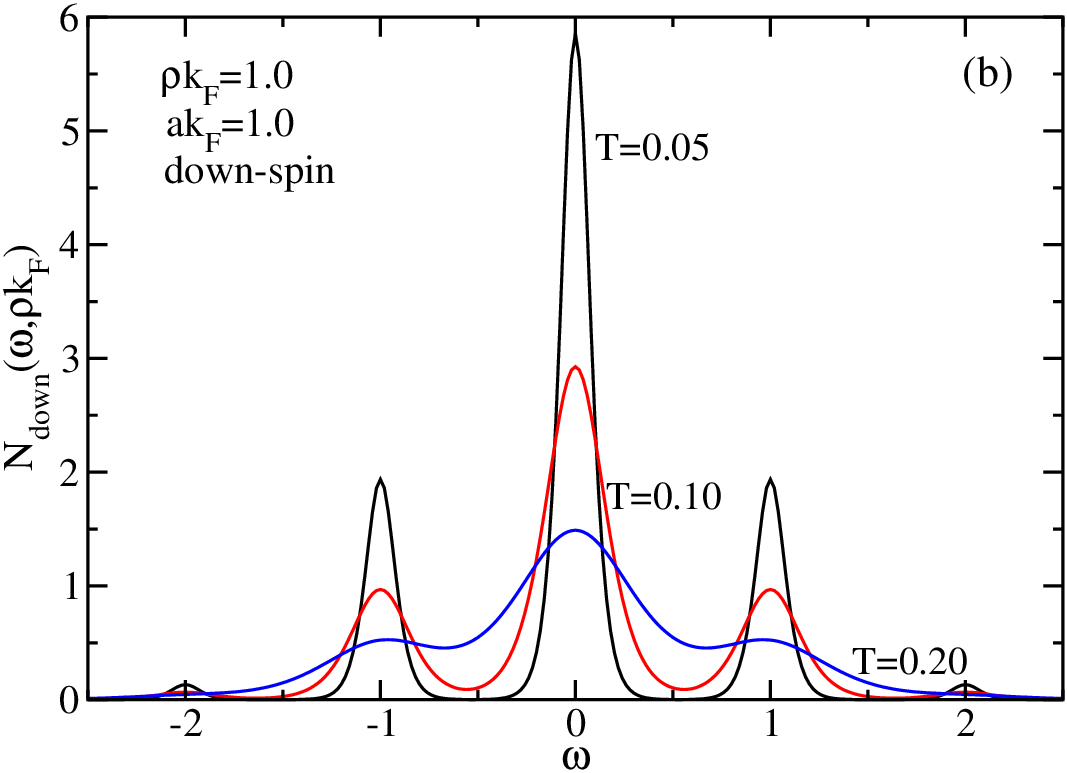}}
\end{center}
\caption{({\it color online}) Spin-polarized LDOS in arbitrary units of the bound states in the linear vortex as a function of energy $\omega$ for $ak_F=1.0$ and ${\tilde \rho}=1$, (a) up-spin and (b) down-spin. Note that the LDOS for down-spin is symmetric, while the LDOS for up-spin is asymmetric, shifted by one unit towards positive energies. The three curves in each panel represent different temperatures: $T=0.05$ (black), $T=0.1$ (red) and $T=0.2$ (blue). The intensity decreases rapidly with the temperature. }
\label{polarization}
\end{figure*}

Of interest is also the spin-polarization of the peaks in the LDOS. For $a \to 0$ (point vortex) and ${\tilde \rho} =0$ the Majorana state is purely a down-spin peak. With increasing $a$ and ${\tilde \rho}$ a growing up-spin component arises. On the other hand, the $m=1$ peak is predominantly up-spin, with growing down-spin components as $a$ and ${\tilde \rho}$ increase. This pattern is shown in Fig. (\ref{polarization}) and can be expressed as $N_{up}(\omega,{\tilde \rho}) = N_{down}(\omega-1,{\tilde \rho})$ with $\omega$ in units of the excitation energy $E_1$. Note that $N_{down}(\omega,{\tilde \rho})$ is a symmetric function of $\omega$, while $N_{up}(\omega,{\tilde \rho})$ is even about $\omega=1$, and the sum of $N_{up}(\omega,{\tilde \rho})$ and $N_{down}(\omega,{\tilde \rho})$ yields the total LDOS, $N(\omega,{\tilde \rho})$, which is symmetric about $\omega = 0.5$.

\begin{figure*}[!ht]
\vskip 0.2in
\begin{center}
\resizebox{0.45\textwidth}{!}
{\includegraphics{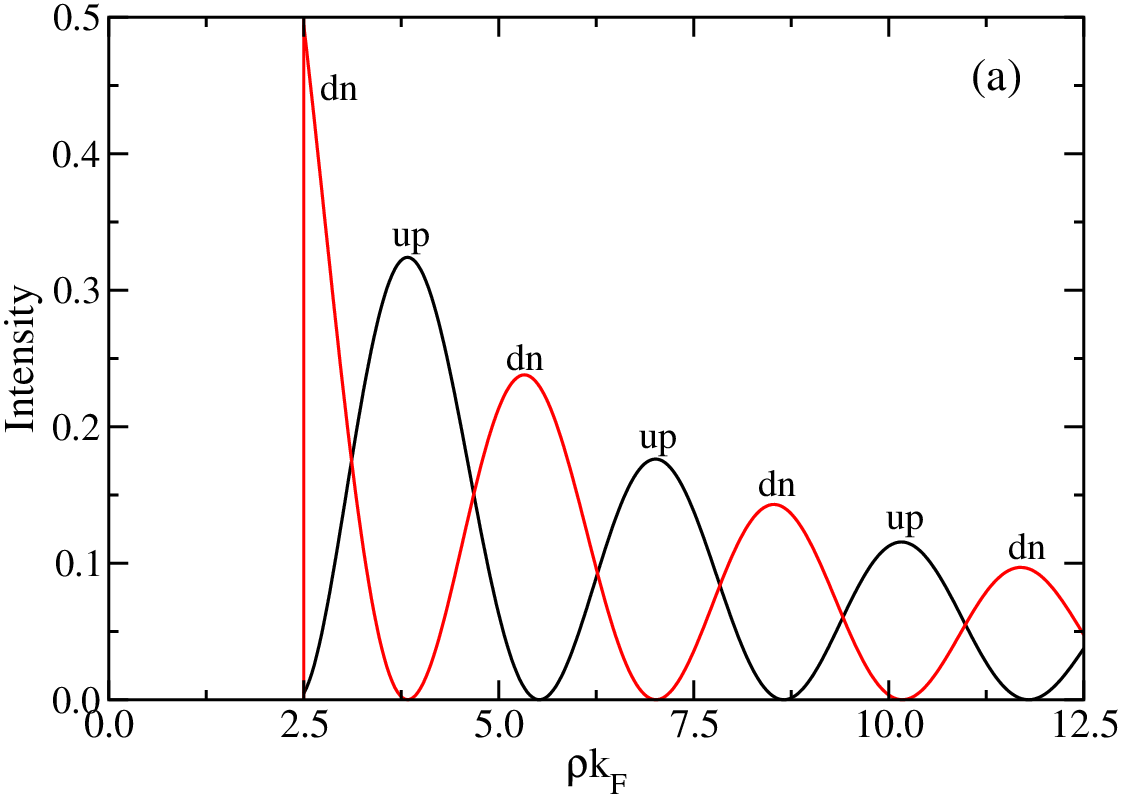}}
\hskip 0.34in
\resizebox{0.45\textwidth}{!}
{\includegraphics{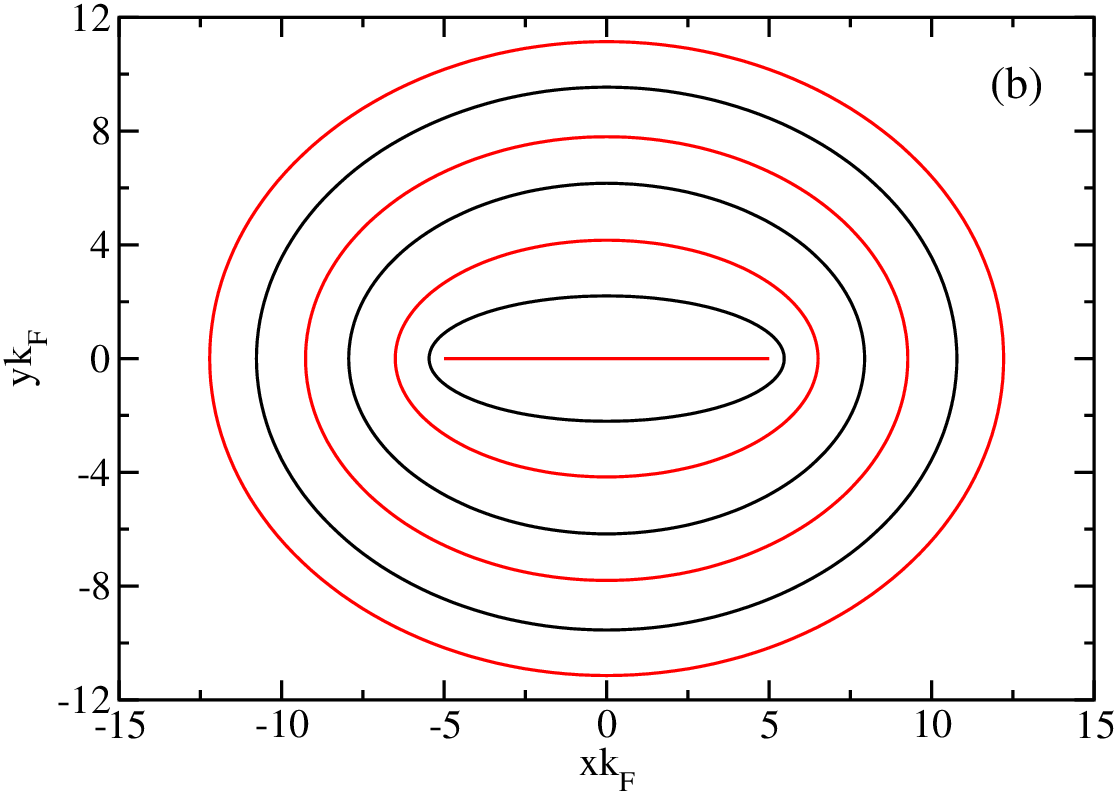}}
\end{center}
\caption{({\it color online}) (a) Spin-polarized maxima of the LDOS in arbitrary units for the Majorana state in the linear vortex as a function of ${\tilde \rho}$ for $ak_F=5.0$. Up- (black) and down-spin (red) maxima alternate and decrease monotonically with ${\tilde \rho}$. (b) The orbital pattern are circles with growing radius for coordinates ${\tilde \rho}$ and $\theta$, while they are ellipses for real space coordinates, $x$ and $y$. The red straight line represents the gapless region.}
\label{polarMajorana}
\end{figure*}

\subsection{Orbits in Cartesian coordinates}

So far we investigated the solution of the bound states in terms of the natural elliptic coordinates, ${\tilde \rho}$ and $\nu$. These variables do not correspond to the real space Cartesian coordinates $x$ and $y$, defined by the elliptic coordinates Eq. (\ref{elliptic1}). We now consider the zero-energy Majorana state, which is given by the zero-order (down-spin density) and first-order (up-spin density) Bessel functions. The oscillations of the Bessel functions give rise to maxima and minima, which are shown in Fig. (\ref{polarMajorana}(a)) as a function of ${\tilde \rho}$ for $ak_F=5.0$. Here down-spin maxima are displayed in red and up-spin maxima in black. Note that solutions only exist for ${\tilde \rho} \ge ak_F/2$. As a function of ${\tilde \rho}$ the down-spin and up-spin maxima alternate and are approximately equally spaced.

Eq. (\ref{elliptic1}) transforms circular solution (natural coordinated for this problem) into elliptical real space coordinates. Note that for ${\tilde \rho} = ak_F/2$ the real space pattern is a horizontal segment between $-a$ and $a$, which corresponds to a vanishing superconducting gap. The order parameter grows linearly away from this segment. The down- and up-spin maxima ellipses alternate and grow with increasing ${\tilde \rho}$ (see Fig. (\ref{polarMajorana}(b))).

\section{Concluding remarks}

We investigated the bound states in the vortex core of a line-shaped cut of length $2a$ in a two-dimensional topological superconductor with Rashba interaction by solving the BdG equations following the analytic method outlined by Caroli {\it et al}. \cite{Caroli1} The electron gas corresponds to the surface states of a 3D TI with proximity induced superconductivity from a nearby $s$ wave superconductor. \cite{FuKane1,Sau2,Chiu,Rakhmanov} This way the model is reduced to an effective 2D superconductor with a linear pinning defect between the two foci. The superconductor is gapless along the linear defect. The strong spin-orbit interaction locks the momentum and the spin perpendicularly. The characteristic energy scale for the spacing of the energy levels in the vortex is proportional to $\Delta_{\infty}^2/E_F$.

The calculation yields a string of fermion bound states with energy $E_m$, $m \neq 0$, and a bound state with Majorana statistics for $m=0$ and $E=0$. In the present case the pinning defect has a linear shape of length $2a$. The results are similar to those of a point defect studied previously, \cite{Schl1,Schl2,Hassler} except that instead of concentric circles the curves of constant energy now are ellipses. The gapless region in the present case is the segment of length $2a$, while for the point defect the gap only vanishes at the singular point. The analytical expressions for the wave functions, consist of products of a Bessel function and an exponential decay, $\exp(-2K({\tilde \rho}))$, as function of the parameter ${\tilde \rho}$ (which for an ellipse is different than the radius of the circle). Given the wave functions we obtained an analytic expression of the LDOS for the bound states. This quantity is experimentally accessible via STM. \cite{Wang}

We investigated the LDOS as a function of external frequency, temperature, the distance from the center of the vortex and the length $2a$ between the foci. The intensity of the peaks in the LDOS rapidly decreases with temperature due to the smearing of the Fermi function. It is interesting to notice that the particle-hole symmetry is broken in the LDOS as a consequence of the spin-orbit coupling. This can be traced to Eq. (\ref{h}).  For instance, the Majorana peak is a pure down-spin state at the core of a point vortex, but becomes partially polarized away from the center and for finite $a$.

The main difference between the ordinary superconductor and the topological superconducting gas is the spin-locking. In the latter in a closed path the spin is forced to follow the momentum giving rise to a non-trivial Berry phase of 1/2. This converts the half-integer quantum numbers into integer ones and opens the possibility to the existence of a Majorana fermion. 

The experimental search for zero-energy modes was successful in two systems, namely, heterostructures of Bi$_2$Te$_3$/NbSe$_2$ \cite{Sun,Xu,SunJia} and monolayers of the high-temperature superconductor FeTe$_{0.55}$Se$_{0.45}$ on SrTiO$_3$(0011) \cite{Wang,Kong,Machida,Song,Chen} Both systems have the advantage of relatively large superconducting transition temperatures. A zero-energy mode, however does not imply a Majorana state, since it may as well originate from a magnetic impurity or an Andreev reflection. The verification that a zero-energy mode corresponds to a Majorana state requiresbraiding, i.e. exchanging the positions of two Majorana zero modes.

Waterfall-like STM spectra have been observed in the iron superconductors \cite{Wang,Chen} and were theoretically studied in \cite{Schl3}. Zero-energy bound states simultaneously appearing at both ends of a 1D atomic line defect were discovered in \cite{Chen}. These line defects resemble the ones studied in this paper.

\appendix \vskip 0.2in
{\bf  \centerline{\large Appendix:} \centerline{Solution of Bogoliubov-de Gennes equations}}
\vskip 0.2in

The method of CdeGM \cite{Caroli1,Caroli2,Schl1} consists of solving the equations (i) for small ${\tilde \rho}$ and (ii) for larger ${\tilde \rho}$, but distances smaller than the coherence length $\xi$. The solutions for the two regimes are then matched at an intermediate distance, ${\tilde \rho}_c$. If the matching condition is independent of ${\tilde \rho}_c$, the solution is valid for the entire vortex region. This condition also determines the bound state energies inside the vortex. 

\subsection{Second order differential equations}

We first convert the first order differential equations, Eqs. (\ref{f1b}) - (\ref{f4b}), into second order differential equations.\cite{Caroli1,Caroli2,Schl1,Schl2} For instance, we express $f^m_2$ from Eq. (\ref{f1b}) and insert it into Eq. (\ref{f2b}) and similar substitutions for $f^m_1$, $f^m_3$ and $f^m_4$. Defining $q_p = 1 +{\tilde E}$ and $q_h = 1 - {\tilde E}$ for particles and holes, respectively, we obtain 
\begin{eqnarray}
&&\left[\frac{\partial^2}{\partial {\tilde \rho}^2} + \frac{1}{\tilde \rho}\frac{\partial}{\partial {\tilde \rho}} - \frac{(m-1)^2}{{\tilde \rho}^2} + q_p^2\right] f^m_1({\tilde \rho}) 
=q_p {\tilde \Delta} f^m_4({\tilde \rho}) - \left( \frac{\partial}{\partial {\rho}} + \frac{m}{\tilde \rho}\right) {\tilde \Delta} f^m_3({\tilde \rho}) \ \ , \label{f1c} \\
&&\left[\frac{\partial^2}{\partial {\tilde \rho}^2} + \frac{1}{\tilde \rho}\frac{\partial}{\partial {\tilde \rho}} - \frac{m^2}{{\tilde \rho}^2} + q_p^2\right] f^m_2({\tilde \rho}) 
=-q_p {\tilde \Delta} f^m_3({\tilde \rho}) - \left( \frac{\partial}{\partial {\rho}} - \frac{m-1}{\tilde \rho}\right) {\tilde \Delta} f^m_4({\tilde \rho}) \ \ , \label{f2c} \\
&&\left[\frac{\partial^2}{\partial {\tilde \rho}^2} + \frac{1}{\tilde \rho}\frac{\partial}{\partial {\tilde \rho}} - \frac{(m+1)^2}{{\tilde \rho}^2} + q_h^2\right] f^m_3({\tilde \rho}) 
=q_h {\tilde \Delta} f^m_2({\tilde \rho}) - \left( \frac{\partial}{\partial {\rho}} - \frac{m}{\tilde \rho}\right) {\tilde \Delta} f^m_1({\tilde \rho}) \ \ , \label{f3c} \\
&&\left[\frac{\partial^2}{\partial {\tilde \rho}^2} + \frac{1}{\tilde \rho}\frac{\partial}{\partial {\tilde \rho}} - \frac{m^2}{{\tilde \rho}^2} + q_h^2\right] f^m_4({\tilde \rho}) 
=-q_h {\tilde \Delta} f^m_1({\tilde \rho}) - \left( \frac{\partial}{\partial {\rho}} + \frac{m+1}{\tilde \rho}\right) {\tilde \Delta} f^m_2({\tilde \rho}) \ \ . \label{f4c}
\end{eqnarray}
 
\subsection{Solution for ${\tilde \rho} < {\tilde \rho}_c$}

Since $2a \ll \xi$, where $\xi$ is the coherence length, for ${\tilde \rho} < {\tilde \rho}_c$ we may neglect ${\tilde \Delta}$ in the core of the vortex. The equations for ${\tilde \rho} < {\tilde \rho}_c$ are then of the integer Bessel function type
\begin{eqnarray}
&&\left[\frac{\partial^2}{\partial {\tilde \rho}^2} + \frac{1}{\tilde \rho}\frac{\partial}{\partial {\tilde \rho}} - \frac{(m-1)^2}{{\tilde \rho}^2} + q_p^2\right] f^m_1({\tilde \rho}) = 0 \ \ , \label{f1d} \\
&&\left[\frac{\partial^2}{\partial {\tilde \rho}^2} + \frac{1}{\tilde \rho}\frac{\partial}{\partial {\tilde \rho}} - \frac{m^2}{{\tilde \rho}^2} + q_p^2\right] f^m_2({\tilde \rho}) = 0 \ \ , \label{f2d} \\
&&\left[\frac{\partial^2}{\partial {\tilde \rho}^2} + \frac{1}{\tilde \rho}\frac{\partial}{\partial {\tilde \rho}} - \frac{(m+1)^2}{{\tilde \rho}^2} + q_h^2\right] f^m_3({\tilde \rho}) = 0 \ \ , \label{f3d} \\
&&\left[\frac{\partial^2}{\partial {\tilde \rho}^2} + \frac{1}{\tilde \rho}\frac{\partial}{\partial {\tilde \rho}} - \frac{m^2}{{\tilde \rho}^2} + q_h^2\right] f^m_4({\tilde \rho}) = 0 \ \ . \label{f4d}
\end{eqnarray}
For $m \ge 1$, the solutions for ${\tilde \rho} < {\tilde \rho}_c$ are then
\begin{eqnarray}
&&f_1^m(q_p{\tilde \rho}) = A_1^m J_{m-1}(q_p {\tilde \rho}) \ , \label{f1e} \\
&&f_2^m(q_p{\tilde \rho}) = A_2^m J_m(q_p {\tilde \rho}) \ , \label{f2e} \\
&&f_3^m(q_h{\tilde \rho}) = A_3^m J_{m+1}(q_h {\tilde \rho}) \ , \label{f3e} \\
&&f_4^m(q_h{\tilde \rho}) = A_4^m J_m(q_h {\tilde \rho}) \ . \label{f4e}
\end{eqnarray}
The constants $A_j^m$ are not all independent. Inserting the solutions into the first order differential equations (still with $\Delta=0$) we obtain $A^m_1 = A^m_2$ and $A^m_3 = -A^m_4$. $A_1^m$ and $A_4^m$ are independent for $\Delta=0$, but are coupled for $\Delta \neq 0$, so that $A_1=\sqrt{1 +{\tilde E}}$ and $A_4 = \sqrt{1 - {\tilde E}}$ (see Appendix A.4).

\subsection{Solution for ${\tilde \rho} > {\tilde \rho}_c$}

In this case $\Delta$ plays an important role. An Ansatz for a solution of the second order differential equations is in terms of a Hankel function of the first kind times an envelope function $g_j$ \cite{Bardeen}
\begin{equation}
f_j^m({\tilde \rho}) = B_j [H^{(1)}_m({\tilde \rho}) g_j({\tilde \rho}) + c.c.] \ , \label{Ansatz}
\end{equation}
where $\rho \ll \xi$ and the $B_j$ are constants. We postulate $B_1=B_2=-B_3=B_4={\textstyle \frac{1}{2}}B$, where the normalization constant $B$ is equated to one for simplicity. The relative phases of the $B_j$ are the same as in Eqs. (\ref{f1e})-(\ref{f4e}). This choice of $B_j$ is verified below where the wave function for ${\tilde \rho} < {\tilde \rho}_c$ is matched to the one for ${\tilde \rho} > {\tilde \rho}_c$.

It is convenient to rewrite Eq. (\ref{f1c}) as
\begin{eqnarray}
&&\left[\frac{\partial^2}{\partial {\tilde \rho}^2} + \frac{1}{\tilde \rho}\frac{\partial}{\partial {\tilde \rho}} - \frac{m^2}{{\tilde \rho}^2} + 1\right] f^m_1({\tilde \rho}) + \left({\tilde E}^2 + 2 {\tilde E} + \frac{2m-1}{{\tilde \rho}^2} \right) f^m_1({\tilde \rho}) \nonumber \\
&&=(1+{\tilde E}) {\tilde \Delta} f^m_4({\tilde \rho}) - \left( \frac{\partial}{\partial {\rho}} + \frac{m}{\tilde \rho}\right) {\tilde \Delta} f^m_3({\tilde \rho}) \ \ , \label{f1f} 
\end{eqnarray}
and similarly for the remaining three equations, Eqs. (\ref{f2c})-(\ref{f4c}). The first line corresponds to the differential equation of the Hankel function. Inserting Eq. (\ref{Ansatz}) into Eq. (\ref{f1f}), dividing by $H_m^{(1)}({\tilde \rho})$ and using the asymptotic expansion of the Hankel function for large argument,
\begin{equation} 
\frac{1}{H^{(1)}_m({\tilde \rho})} \frac{d H^{(1)}_m({\tilde \rho})}{d {\tilde \rho}} \approx -\frac{1}{2 {\tilde \rho}} +i \ . \label{asymt}
\end{equation}
we obtain the following differential equations for the functions $g_j({\tilde \rho})$:
\begin{eqnarray}
\frac{d^2 g_1}{d {\tilde \rho}^2} &+& 2i \frac{d g_1}{d{\tilde \rho}} + \Bigl[{\tilde E}^2 + 2{\tilde E} + \frac{2m-1}{{\tilde \rho}^2} \Bigr] g_1 \nonumber \\
&=& (1+{\tilde E}) {\tilde \Delta} g_4 + {\tilde \Delta} \frac{d g_3}{d {\tilde \rho}} + \frac{d{\tilde \Delta}}{d{\tilde \rho}} g_3 + \Bigl[\frac{2m-1}{2{\tilde \rho}} + i\Bigr] {\tilde \Delta} g_3  \ \ , \label{f1g} \\
\frac{d^2 g_2}{d {\tilde \rho}^2} &+& 2i \frac{d g_2}{d{\tilde \rho}} + \Bigl[{\tilde E}^2 + 2{\tilde E} \Bigr] g_2 = (1+{\tilde E}) {\tilde \Delta} g_3\nonumber \\
&-& {\tilde \Delta} \frac{d g_4}{d {\tilde \rho}} + \Bigl[\frac{2m-1}{2{\tilde \rho}} - i\Bigr] {\tilde \Delta} g_4 - \frac{d{\tilde \Delta}}{d{\tilde \rho}} g_4 \ \ , \label{f2g} \\
\frac{d^2 g_3}{d {\tilde \rho}^2} &+& 2i \frac{d g_3}{d{\tilde \rho}} + \Bigl[{\tilde E}^2 - 2{\tilde E} - \frac{2m+1}{{\tilde \rho}^2} \Bigr] g_3 \nonumber \\
&=& -(1-{\tilde E}) {\tilde \Delta} g_2 + {\tilde \Delta} \frac{d g_1}{d {\tilde \rho}} + \frac{d{\tilde \Delta}}{d{\tilde \rho}} g_1 - \Bigl[\frac{2m+1}{2{\tilde \rho}} - i\Bigr] {\tilde \Delta} g_1     \ \ , \label{f3g} \\
\frac{d^2 g_4}{d {\tilde \rho}^2} &+& 2i \frac{d g_4}{d{\tilde \rho}} + \Bigl[{\tilde E}^2 - 2{\tilde E}\Bigr] g_4 = -(1-{\tilde E}) {\tilde \Delta} g_1\nonumber \\
&-& {\tilde \Delta} \frac{d g_2}{d {\tilde \rho}} - \bigl[\frac{2m+1}{2{\tilde \rho}} + i\Bigr] {\tilde \Delta} g_2 - \frac{d{\tilde \Delta}}{d{\tilde \rho}} g_2 \ \ . \label{f4g} 
\end{eqnarray}

These equations are solved iteratively. To zeroth order we keep the dominant terms,
\begin{eqnarray}
2i\frac{d}{d {\tilde \rho}} \left[ \begin{array}{c} g_1^{(0)} \\ g_2^{(0)} \\ g_3^{(0)} \\ g_4^{(0)} \end{array} \right] = {\tilde \Delta} \left[ \begin{array}{c} g_4^{(0)} \\ g_3^{(0)} \\ -g_2^{(0)} \\ -g_1^{(0)} \end{array} \right] + i {\tilde \Delta} \left[ \begin{array}{c} g_3^{(0)} \\ -g_4^{(0)} \\ g_1^{(0)} \\ -g_2^{(0)} \end{array} \right] \ . \label{g0}
\end{eqnarray}
The remaining terms are first order terms and will be treated perturbatively. The solution of Eq. (\ref{g0}) is of the form
\begin{eqnarray}
&&g^{(0)}_1({\tilde \rho}) = C e^{-K({\tilde \rho})} , \ \ \ \ \ g^{(0)}_2({\tilde \rho}) = -iC e^{-K({\tilde \rho})} , \nonumber \\
&&g^{(0)}_3({\tilde \rho}) = - C e^{-K({\tilde \rho})} , \ \ \ g^{(0)}_4({\tilde \rho}) = -iC e^{-K({\tilde \rho})} , \label{g1-4}
\end{eqnarray}
where $K({\tilde \rho})$ has been defined previously in Eq. (\ref{K}) and $C$ is a unitary constant $e^{i\gamma}$ to be determined later.

The equations for the first order terms in Eqs. (\ref{f1g})-(\ref{f4g}) are
\begin{eqnarray}
&&2i\frac{d}{d {\tilde \rho}} \left[ \begin{array}{c} g_1^{(1)} \\ g_2^{(1)} \\ g_3^{(1)} \\ g_4^{(1)} \end{array} \right] - {\tilde \Delta} \left[ \begin{array}{c} g_4^{(1)} \\ g_3^{(1)} \\ -g_2^{(1)} \\ -g_1^{(1)} \end{array} \right] - i {\tilde \Delta} \left[ \begin{array}{c} g_3^{(1)} \\ -g_4^{(1)} \\ g_1^{(1)} \\ -g_2^{(1)} \end{array} \right] =-\frac{d^2}{d {\tilde \rho}^2} \left[ \begin{array}{c} g_1^{(0)} \\ g_2^{(0)} \\ g_3^{(0)} \\ g_4^{(0)} \end{array} \right] \nonumber \\
&&- \left[\begin{array}{c} ({\tilde E}^2 + 2 {\tilde E} + \frac{2m-1}{{\tilde \rho}^2})g_1^{(0)}  \\ ({\tilde E}^2 + 2 {\tilde E} ) g_2^{(0)} \\  ({\tilde E}^2 - 2 {\tilde E} - \frac{2m+1}{{\tilde \rho}^2})g_3^{(0)}  \\ ({\tilde E}^2 - 2 {\tilde E} ) g_4^{(0)} \end{array}\right] +{\tilde \Delta} {\tilde E} \left[ \begin{array}{c} g_4^{(0)} \\ g_3^{(0)} \\ g_2^{(0)} \\ g_1^{(0)} \end{array} \right] + {\tilde \Delta} \frac{d}{d {\tilde \rho}} \left[ \begin{array}{c} g_3^{(0)}  \\  -g_4^{(0)} \\ g_1^{(0)} \\ -g_2^{(0)} \end{array}\right] \nonumber \\
&&+\frac{d{\tilde \Delta}}{d {\tilde \rho}} \left[ \begin{array}{c} g_3^{(0)}  \\  -g_4^{(0)} \\ g_1^{(0)} \\ -g_2^{(0)} \end{array}\right] +\frac{{\tilde \Delta}}{2{\tilde \rho}} \left[ \begin{array}{c} (2m-1)g_3^{(0)}  \\  (2m-1)g_4^{(0)} \\ -(2m+1)g_1^{(0)} \\ -(2m+1)g_2^{(0)} \end{array}\right] \ . \label{firstorder}
\end{eqnarray}
Inserting the zero-order solutions, Eq. (\ref{g1-4}), into Eq. (\ref{firstorder}) and considering the Ansatz $g^{(1)}_1 = C a_1 e^{-K}$, $g^{(1)}_2 = -i C a_2 e^{-K}$, $g^{(1)}_3 = -C a_3 e^{-K}$ and $g^{(1)}_4 = -i C a_4 e^{-K}$, we obtain
\begin{eqnarray}
&&2i\frac{d}{d {\tilde \rho}} \left[ \begin{array}{c} a_1 \\ -ia_2 \\ -a_3 \\ -ia_4 \end{array} \right] - 2i{\tilde \Delta} \left[ \begin{array}{c}  a_1 \\ -ia_2 \\ -a_3 \\ -ia_4 \end{array} \right] -{\tilde \Delta} \left[ \begin{array}{c} -ia_4 \\ -a_3 \\ ia_2 \\ -a_1 \end{array} \right] - i {\tilde \Delta} \left[ \begin{array}{c} -a_3 \\ ia_4 \\ a_1 \\ ia_2 \end{array} \right] \nonumber \\
&&= -\left[\begin{array}{c} ({\tilde E}^2 + 2 {\tilde E} + \frac{2m-1}{{\tilde \rho}^2})  \\ -i({\tilde E}^2 + 2 {\tilde E} )  \\  -({\tilde E}^2 - 2 {\tilde E} - \frac{2m+1}{{\tilde \rho}^2}) \\ -i({\tilde E}^2 - 2 {\tilde E} ) \end{array}\right] +{\tilde \Delta} {\tilde E} \left[ \begin{array}{c} -i \\ -1 \\ -i \\ 1 \end{array} \right] 
+\frac{\tilde \Delta}{2{\tilde \rho}} \left[ \begin{array}{c} -(2m-1)  \\  -i(2m-1) \\ -(2m+1) \\ i(2m+1) \end{array}\right] \ , \label{aj}
\end{eqnarray}
where terms of the order $\frac{d {\tilde \Delta}}{d{\tilde \rho}}$ and ${\tilde \Delta}^2$ vanish identically. All terms are proportional to $e^{-K{\tilde \rho}}$, so that this factor cancels out. The differential equations then take the following form
\begin{eqnarray}
&&2\frac{d}{d {\tilde \rho}} \left[ \begin{array}{c} a_1 \\ a_2 \\ a_3 \\ a_4 \end{array} \right] + {\tilde \Delta} \left[ \begin{array}{c}  (a_3+a_4)-2a_1 \\ (a_3+a_4)-2a_2 \\ (a_1+a_2)-2a_3 \\ (a_1+a_2)-2a_4 \end{array} \right] = -{\tilde \Delta} {\tilde E} \left[ \begin{array}{c} 1 \\ 1 \\ -1 \\ -1 \end{array} \right] \nonumber \\      
&&+i\left[\begin{array}{c} {\tilde E}^2 + 2 {\tilde E} + \frac{2m-1}{{\tilde \rho}^2}  \\ {\tilde E}^2 + 2 {\tilde E}  \\  {\tilde E}^2 - 2 {\tilde E} - \frac{2m+1}{{\tilde \rho}^2} \\ {\tilde E}^2 - 2 {\tilde E}  \end{array}\right] +i\frac{\tilde \Delta}{2{\tilde \rho}} \left[ \begin{array}{c} 2m-1  \\  -2m-1 \\ -2m+1 \\ 2m+1 \end{array}\right] \ , \label{ajx}
\end{eqnarray}
which can be decoupled by taking linear combinations,
\begin{eqnarray}
&&2\frac{d}{d {\tilde \rho}} (a_1-a_2) - 2 {\tilde \Delta}(a_1-a_2) = i(2m-1)\Bigl[\frac{1}{{\tilde \rho}^2} + \frac{{\tilde \Delta}}{{\tilde \rho}}\Bigr] \nonumber \\ 
&&2\frac{d}{d {\tilde \rho}} (a_3-a_4) - 2 {\tilde \Delta}(a_3-a_4) = -i(2m+1)\Bigl[\frac{1}{{\tilde \rho}^2} + \frac{{\tilde \Delta}}{{\tilde \rho}}\Bigr] \nonumber \\
&&\frac{d}{d{\tilde \rho}}(a_1+a_2+a_3+a_4) = 2i{\tilde E}^2 - i \frac{1}{{\tilde \rho}^2} \nonumber \\      
&&\frac{d}{d{\tilde \rho}}(a_1+a_2-a_3-a_4)-2{\tilde \Delta}(a_1+a_2-a_3-a_4) = 4i {\tilde E}+2i\frac{m}{{\tilde \rho}^2} - 2 {\tilde \Delta }{\tilde E} . \label{ajx1}
\end{eqnarray}

The above equations can be integrated
\begin{eqnarray}
a_1-a_2 &=& - i(m-{\textstyle \frac{1}{2}}) \int_{\tilde \rho}^{\infty} dx \exp[K({\tilde \rho}) - K(x)] \Bigl[\frac{1}{x^2} + \frac{{\tilde \Delta}}{x}\Bigr] \label{int1} \\ 
a_3-a_4 &=& i(m+{\textstyle \frac{1}{2}}) \int_{\tilde \rho}^{\infty} dx \exp[K({\tilde \rho}) - K(x)] \Bigl[\frac{1}{x^2} + \frac{{\tilde \Delta}}{x}\Bigr] \label{int2} \\ 
a_1+a_2 &+& a_3+a_4 = 2i{\tilde E}^2 {\tilde \rho} + \frac{i}{\tilde \rho} \label{int3} \\ 
a_1+a_2 &-& a_3-a_4 = - i \int_{\tilde \rho}^{\infty} dx \exp[2K({\tilde \rho}) - 2K(x)] \Bigl[4{\tilde E} + \frac{2m}{x^2} \Bigr] \nonumber \\
&-&2\int_0^{\tilde \rho}dx \exp[2K({\tilde \rho}) - 2K(x)] {\tilde E} {\tilde \Delta} . \label{int4}
\end{eqnarray}
We are only interested in leading and next-leading terms and will disregard higher order terms. Small parameters are ${\tilde \Delta}{(\tilde \rho)}$, ${\tilde E}$ and ${\tilde \rho}$, so that the term ${\tilde E}^2 {\tilde \rho}$ of Eq. (\ref{int3}) and the last term of Eq. (\ref{int4}) can be neglected.

The remaining two integrals can be integrated by parts:
\begin{equation}
\int_{\tilde \rho}^{\infty}dx \exp[K({\tilde \rho})-K(x)] \Bigl(\frac{1}{x^2} + \frac{\tilde \Delta}{x} \Bigr) = \frac{1}{\tilde \rho} \ , \label{byparts1}
\end{equation}
and
\begin{eqnarray}
&&\int_{\tilde \rho}^{\infty} dx\exp[2 K({\tilde \rho}) -2 K(x)] \Bigl( {\tilde E}+\frac{m}{2x^2}\Bigr) = -{\tilde E}{\tilde \rho} +\frac{m}{2 {\tilde \rho}}+ G({\tilde \rho}) \label{byparts2} \\
&&G({\tilde \rho}) = \int^{\infty}_{\tilde \rho} dx \exp[2 K({\tilde \rho}) -2 K(x)] 2{\tilde \Delta}(x) \Bigl({\tilde E}x-\frac{m}{2x}\Bigr) .  \label{G} 
\end{eqnarray}
The function $G({\tilde \rho})$ equals zero at the matching point (see Appendix A.4) and defines the bound state energies.

It is now straightforward to solve Eqs. (\ref{int1})-(\ref{int4}) for $a_j$
\begin{eqnarray}
&&a_1 = i {\tilde E}{\tilde \rho} - i \frac{2m-1}{2 {\tilde \rho}} -i G({\tilde \rho}) \nonumber \\
&&a_2 = i {\tilde E}{\tilde \rho} -i G({\tilde \rho}) \nonumber \\
&&a_3 = -i {\tilde E}{\tilde \rho} + i \frac{2m+1}{2 {\tilde \rho}} +i G({\tilde \rho}) \nonumber \\
&&a_4 = -i {\tilde E}{\tilde \rho} +i G({\tilde \rho}) \label{ajj} \ .
\end{eqnarray}
This corresponds to the leading and next-leading solution for the wave function for large argument.

\subsection{Matching of wave functions}

The next step consists in matching the solution for the wave function for small ${\tilde \rho}$ and large ${\tilde \rho}$ at an intermediate value ${\tilde \rho}_c$ from 
the core of the vortex. The wave function is determined uniquely if this matching is independent of ${\tilde \rho}_c$. This condition on ${\tilde \rho}_c$ also implies that $G({\tilde \rho}_c) \approx 0$ and consequently the energy for the $m^{th}$ bound state is given by
\begin{equation}
E_m \int_{{\tilde \rho}_c}^{\infty} dx \exp[-2K(x)] = m \int_{{\tilde \rho}_c}^{\infty} dx \exp[-2K(x)] \frac{\tilde \Delta}{x} . \label{Em} 
\end{equation}
The excited states are then equally spaced energy levels.

The solutions for ${\tilde \rho} > {\tilde \rho}_c$ are given by $f^m_j({\tilde \rho}) = B_j [H_m^{(1)}({\tilde \rho}) g_j({\tilde \rho}) +c.c.]$, where the coefficients $B_j$ are defined just after (Eq. (\ref{Ansatz})). The functions $g_j$ were calculated consistently to first order of perturbation and can be written as an exponential, i.e. $g_j({\tilde \rho}) \propto \exp[-K({\tilde \rho}) + a_j({\tilde \rho})]$. This expression remains correct to first order. For ${\tilde \rho} < {\tilde \rho}_c$ the solutions are Eqs. (\ref{f1e}) - (\ref{f4e}), which have to be matched at ${\tilde \rho}_c$ following a procedure similar to Refs. \cite{Schl1} and \cite{Caroli1}. To match the wave functions we use the asymptotic expansions for the Bessel and Hankel functions. To simplify we only explicitly work out the matching for the function $f_1^m$; the remaining three functions follow similarly. 

For ${\tilde \rho} < {\tilde \rho}_c$ we have
\begin{eqnarray}
f_1^m({\tilde \rho}_c) &&= A_1 J_{m-1}\bigl((1+{\tilde E}){\tilde \rho}_c\bigr) \exp(-K({\tilde \rho}_c)) \nonumber \\
&&=\sqrt{\frac{2}{\pi{\tilde \rho}_c}}\cos\Bigl[(1+{\tilde E}){\tilde \rho}_c -\frac{\pi}{2}(m-1) -\frac{\pi}{4} \nonumber \\
&&+\frac{(m-1)^2 - \frac{1}{4}}{2(1+{\tilde E}){\tilde \rho}_c }\Bigr] \exp(-K({\tilde \rho}_c)) , \label{small}
\end{eqnarray}
and similarly for ${\tilde \rho} > {\tilde \rho}_c$ we obtain (with $B_1=\frac{1}{2}$)
\begin{eqnarray}
f_1^m({\tilde \rho}_c) &&= \frac{1}{2} \Bigl[H^{(1)}_m({\tilde \rho}_c) g_1({\tilde \rho}_c) + c.c.\Bigr] B_1 \nonumber \\
&&=\sqrt{\frac{2}{\pi {\tilde \rho}_c}} \Bigl\{\exp\Bigl[i\bigl({\tilde \rho}_c -\frac{\pi m}{2} -\frac{\pi}{4} +\frac{m^2-\frac{1}{4}}{2 {\tilde \rho}_c}\bigr) 
\Bigr] \nonumber \\
&&\times\exp\Bigl[i\bigl(\gamma +{\tilde E}{\tilde \rho}_c-\frac{2m-1}{2{\tilde \rho}_c}\bigr)\Bigr] + c.c. \Bigr\} \frac{1}{2} \exp(-K({\tilde \rho}_c)) . \label{large}
\end{eqnarray}
The phase $\gamma$ arises from the constant $C$ for the solution for ${\tilde \rho} > {\tilde \rho}_c$. For $\gamma = \frac{\pi}{2}$ the two expressions, (\ref{small}) and (\ref{large}), are identical to next-leading order for all four functions, $j=1,\cdots,4$.
\vskip 0.3in
{\centerline {\bf DATA AVAILABILITY}}
\vskip 0.1in
The data that support the findings of this article are not publicly available.  The data are available from the author upon reasonable request.

\end{document}